\documentclass[twocolumn]{aastex631}

\newcommand{\reply}[1]{{#1}}

\usepackage{xcolor}
\usepackage{savesym}
\savesymbol{tablenum}
\usepackage{siunitx}
\restoresymbol{SIX}{tablenum}
\usepackage{float}
\begin{document}

\title{Tracking the Chemical Evolution of Hydrocarbons Through Carbon Grain \reply{Supply} in Protoplanetary Disks}

\author[0009-0002-2380-6683]{Eshan Raul}
\affiliation{Department of Astronomy, University of Michigan, 1085 South University Ave., Ann Arbor, MI 48109, USA}

\author[0000-0002-2692-7862]{Felipe Alarcón}
\affiliation{Department of Astronomy, University of Michigan, 1085 South University Ave., Ann Arbor, MI 48109, USA}
\affiliation{Dipartimento di Fisica, Università degli Studi di Milano, Via Celoria 16, 20133 Milano, Italy}

\author[0000-0003-4179-6394]{Edwin A. Bergin}
\affiliation{Department of Astronomy, University of Michigan, 1085 South University Ave., Ann Arbor, MI 48109, USA}

\begin{abstract}
The gas present in planet-forming disks typically exhibits strong emission features of abundant carbon and oxygen molecular carriers. In some instances, protoplanetary disks show an elevated C/O ratio above interstellar values, which leads to a rich hydrocarbon chemistry evidenced in the mid-infrared spectra. The origin of this strengthened C/O ratio may stem from the release of less complex hydrocarbons from the chemical processing of carbonaceous grains. We have explored a set of \reply{42} single-cell models in which we match the physical conditions to the inner regions of planet-forming disks, while varying the \reply{C/O ratio} by exploring different \reply{levels of CH$_4$, C, H$_2$O, and CO} to the gas-phase chemistry, which we evaluate in \reply{both the cosmic/X-ray and UV-driven limit}. \reply{We find that the carbon-bearing species in our models exhibit high dependencies on the driver of the chemistry, where both CO and long chain hydrocarbons act as carbon sinks in the cosmic/X-ray-driven chemistry limit, while the vast majority ends up in atomic carbon and CO in the UV-driven limit. We also find moderate dependencies upon the C/O ratio, where this and the ionization rate/UV field determines the point of peak production of a species as well as its equilibrium abundance. We also find that the production of several hydrocarbons, specifically C$_2$H$_2$, is strongly dependent up to an order of magnitude on the initial water abundance. We lastly find that in the X-ray-driven limit, both CH$_4$ and C serve as highly transient donor species to the carbon chemistry.}
\end{abstract}
\keywords{astrochemistry --- protoplanetary disks}

\section{Introduction} \label{sec:intro}
Inner solar system solids (e.g., the Earth, meteorites) are extremely carbon deficient in comparison to the carbon content anticipated within the inner few au \citep[][]{Bergin2015}. This is exacerbated by the fact that nearly 50\% of carbon in the interstellar medium (ISM) appears to be in some refractory form \citep{Mishra05, Jones13, Chiar13}. This refractory carbonaceous material appears to be present in comets \citep{Fomenkova1999, Rubin2019}, but is not present even in the most primitive CI chondrites \citep{Bergin2015}. The Earth and smaller rocky bodies should theoretically, without any chemical processing, have a refractory composition similar to that of ISM \citep{Binkert2023}. To limit the supply of carbon-rich solids to terrestrial worlds, some process must exist that destroys ISM refractory carbon within the inner solar system to place this material in the gaseous state. Possibilities include oxidation, photoablation, and thermal processes such as sublimation and pyrolysis \citep{Gail2002, Lee2010, Gail2017, Anderson..2017, Klarmann2018, Li2021}.

Recent observations from the James Webb Space Telescope (JWST) have shown that some disks surrounding young protostars exhibit rich carbon chemistry \citep[e.g.][]{Tabone2023,Arabhavi2024, Kanwar2024}. This chemistry is potentially consistent with an elevated gaseous C/O ratio and the release of hydrocarbons from refractory material via sublimation \citep{Li2021}. If true, this would hint that the formation processes that produce the Earth are replicated elsewhere, and that planets forming in these systems would also be carbon-poor \citep{Tabone2023}.
However, \reply{additional carbon supply terms are theorized to be potentially present.} Thus, it is unclear if the observed chemistry is consistent with the destruction of refractory material\reply{; for instance, the supply of CH$_4$ vapor would potentially be sufficient to trigger this elevated C/O chemistry \citep{Mah2023}. } \reply{Further, this chemistry is observed predominantly towards very low mass (M$_\star <$~0.2~M$_\odot$) stars 
\citep{Pascucci2009, Pascucci2013}.  This conflicts with a model that depends on sublimation, as the sublimation fronts should be found at larger distances towards the solar mass disk systems and a carbon-rich chemistry should have been detected in these disks.  }

This hydrocarbon-producing chemistry naturally prompts the question of what it looks like under different inner-region physical conditions \citep[see, e.g.,][]{Wei2019}.   
With the ongoing observations of the JWST, we can expect an abundance of forthcoming discoveries of small organic molecules within disks in the near future. This makes investigation of the production of such species essential in comprehending the chemistry in the inner regions of disks where terrestrial planets may be formed \citep{Dishoeck2023, Kamp2023}.  

Recent JWST observations have also found detectable levels of CH$_3^+$ in an irradiated disk in the Orion Bar \citep{Berne2023}. This molecule lies at the base of interstellar carbon chemistry \citep{Herbst21} and hints that, in some instances, hydrocarbon chemistry is potentially a central aspect of disk evolution. Strong and consistent detections of combinations of H$_2$O, C$_2$H$_2$, HCN, and CO$_2$  in planet-forming disks \citep{Carr2008, Pontoppidan2009, Salyk2011,Kamp2023,Gasman2023,Dishoeck2023,Tannus2023} incentivize an examination of the reactions that lead to the production of those observed molecules (C$_2$H$_2$, HCN, and CO$_2$), and the role that H$_2$O plays within it \citep{Bosman2022, Duval22}. \reply{Within the very low mass  systems the extensive hydrocarbon-dominated chemistry goes well beyond C$_2$H$_2$ with detected emission from C$_4$H$_2$, C$_3$H$_4$, C$_2$H$_6$, C$_6$H$_6$ and CH$_3$ \citep{Tabone2023, Arabhavi2024, Kanwar2024}.} The detection of high C$_2$H$_2$/CO$_2$ and C$_2$H$_2$/H$_2$O column densities in these systems implies there are varying conditions under which C$_2$H$_2$ forms, a constraint of which can lead us to a better understanding of the chemical budget in the planet-forming regions of similar disks. \reply{Analyses of hydrocarbon-rich sources are suggesting that a central parameter in activating the observed chemistry is an elevated C/O ratio in the gas \citep{Kanwar2023, Kanwar2024}.}

While previous studies have focused on the disk chemistry in outer regions, or the effect of planet-disk interactions \citep{Facchini..2018, Nienke..2018, Wei2019, Alarcon2020,  Alarcon2021, Broome..2013}, the present work centers on the varying chemistry at different conditions of ionization rate    \reply{and primary ionization driver (UV or X-ray/Cosmic-ray), along with the overall C/O ratio.}
Our goal is to determine under which conditions hydrocarbons of interest, \reply{such as C$_2$H$_2$, CH$_4$, C$_4$H$_2$, C$_6$H$_6$, and others}, are most prone to forming and reaching equilibrium at high abundance values. This will help provide a pathfinder to elucidate key chemical pathways and potentially isolate \reply{how} excess carbon \reply{is supplied to} disk gas. In this exploration we focus on the ionization effects of X-rays and cosmic-rays on the overall chemical equilibrium \reply{which is placed alongside the impact of UV photons.} X-rays can be an important driver for inner disk chemistry in the UV-shielded layers of the disk surface \citep{Glassgold1997}.  \reply{Concurrently,} ultraviolet (UV) photons are also a key chemical driver \citep{Nomura2005, Kamp2010, Gorti2011}.

In this work, we utilize single-point chemical models with a range of physical conditions clustered by the ionization rate. We have also explored multiple \reply{chemical scenarios with a range of C/O ratios to explore the outcomes of carbon-supply} within an overall context of water-rich and water-poor gas. \reply{The excess carbon supplied could be the result of pebble drift with subsequent carbon grain destruction \citep{Li2021}, possibly the photoablation of small carbon grains \citep{Anderson21},  the advection of carbon-rich gas from the outer disk \citep{Mah2023}, or by an unknown mechanism.} In all, we present simulations of \reply{80} individual models of the time-dependent chemistry. This framework is presented in \S~\ref{sec:method}. In \S~\ref{sec:result}, the models are synthesized to determine what range of conditions, both chemical and physical, favor the formation of key observable species (e.g., \reply{C$_2$H$_2$, CH$_4$, and CO$_2$}) finding commonalities in their abundance values through \reply{hierarchically-sorted time-dependent plots}.
In \S~\ref{sec:discussion}, our discussion includes comparisons to recent literature, an examination of the evolution of carbon in our system, and the implications of water for future disk observations. Our findings are summarized in \S~\ref{sec:conclusion}.

\section{Methodology} \label{sec:method}

\begin{table*}[!ht]
\begin{center}
\caption{Solar Model Initial Chemical Conditions}
\hrule \kern .625mm
\hrule
\label{tab:table1}
\begin{tabular}{cc}
\reply{} & \reply{\hbox{Relative}}\\
\reply{Species} & \reply{\hbox{Abundance}}\\
\hline
H$_2$		& 0.5\\
He		& 0.0975\\
O		& $10^{-14}$\\
N		& $10^{-14}$\\
C$^+$		& $10^{-14}$\\
S		& 1.35$\times 10^{-7}$\\
Si$^+$		& $10^{-9}$\\
Na$^+$		& $10^{-14}$\\
Mg$^+$		& $10^{-14}$\\
Fe$^+$		& $10^{-14}$\\
P$^+$		& $10^{-14}$\\
\end{tabular}
\begin{tabular}{cc}
\reply{} & \reply{\hbox{Relative}}\\
\reply{Species} & \reply{\hbox{Abundance}}\\
\hline
F		& $10^{-14}$\\
Cl$^+$		& $10^{-14}$\\
CH$_4$		& $10^{-14}$\\
CH$_2$		& $10^{-14}$\\
H$_2$O	& 2.25$\times 10^{-4}$\\
CO		& \num{1.5e-4}\\
N$_2$		& $10^{-5}$\\
CH$_3$		& $10^{-14}$\\
CN		& $10^{-14}$\\
NH		& $10^{-14}$\\
e$^-$		& $10^{-9}$\\
\end{tabular}
\end{center}
\end{table*}

In our objective to probe these recent observations, we use a C/O and driver-varying approach to determine which parameter sets lead to a strong production of hydrocarbons. \reply{We compare the equilibrium abundance of small hydrocarbons, some of them typically observed to be present in protoplanetary disks by JWST.}

We explore a grid of physical parameters and C/O ratios in a single-cell model including an extensive set of chemical reactions which are detailed below in \S~\ref{subsec:chem}. \S~\ref{subsec:parameters} first describes the base chemical conditions and physical parameters we use, \S~\ref{subsec:setup} addresses the assumptions made by the model and the nuances of the code, followed by how we determine the reaction pathways in \S~\ref{subsec:reactions}.

\subsection{Chemical Solver} \label{subsec:chem}

The chemical kinetics evolution is done by solving single-cell models. The chemical network is solved using the ODE \texttt{QNDF()} solver \citep{Rackauckas}, which is particularly efficient with large stiff ODE problems such as chemical kinetics.

The code includes all the gas-phase reactions in the UMIST 2012 Database for Astrochemistry \citep{McElroy2013}. There are multiple reaction types included within the UMIST 2012 network \citep{McElroy2013}: two-body reactions, charge-exchange reactions, photochemical reactions, cosmic-rays induced reactions, and radiative photoreactions. We include 536 species for a total of 7349 reactions. \reply{We note that the UMIST 2012 network has recently been updated to a 2022 version \citep{Millar2024}, which we seek to explore the application of our approach in a future work, providing a useful comparison between the two chemical networks}.

\subsubsection{Photochemical Reactions}

For photochemical reactions, the \reply{rate coefficients} are calculated using:

\begin{equation}\label{eq:photochemical}
    k(A_V) = \frac{F_{UV}}{G_{\odot}}\times \alpha \exp{\big(-\gamma A_V \big)} \ \mathrm{s}^{-1},
\end{equation}

\noindent where $\alpha$ and $\gamma$ are reaction constants from the chemical network and $A_V$ is the visual extinction, G$_{\odot}$ the standard ISM UV flux and $F_{UV}$ the incident UV flux. We assume  $F_{UV}$=45G$_{\odot}$ and vary the extinction accordingly to replicate different levels of depth, which results in the equivalent UV fluxes listed in Table \ref{tab:table2}.

\subsubsection{Two-body reaction rates}

The generic two-body reaction rate coefficients are calculated using the modified Arrhenius equation:

\begin{equation}\label{eq:arrhenius}
    k(T) = \alpha \Big(\frac{T}{300} \Big)^{\beta} \exp{ \Big( \frac{-\gamma}{T}\Big) } \ \mathrm{cm^{3} \ s}^{-1},
\end{equation}

\noindent where $\alpha$, $\beta$, and $\gamma$ are constants for each reaction and $T$ is the temperature. 

\subsubsection{Cosmic-ray reactions}

For directly \reply{ionizing} reactions the rate coefficient is:

\begin{equation}\label{eq:ionization}
    k(\zeta) =\Big(\frac{\zeta}{1.36 \times 10^{-17} \ \mathrm{s}^{-1}} \Big)\alpha \ \mathrm{s}^{-1},
\end{equation}

\noindent where $\zeta$ is the ionization rate.

Photo-induced cosmic-ray ionization reactions are also taken into account by the network, which take the form:

\begin{equation}\label{eq:ionization2}
    k(\zeta) = \Big(\frac{\zeta}{1.36 \times 10^{-17} \ \mathrm{s}^{-1}} \Big) \alpha \Big(\frac{T}{300} \Big)^{\beta} \Big( \frac{\gamma}{1 - \omega}\Big) \ \mathrm{s}^{-1},
\end{equation}

\noindent where $\omega$ corresponds to the albedo of
dust grains to dissociative photons, and $\alpha$, $\beta$, and $\gamma$ are reaction constants. In our models, we set $\omega$=0.6. 

\subsubsection{Adsorption and thermal desorption}

The chemical solver also considers gas-grain reactions using the \citep{Hasegawa..92} prescription, including thermal desorption and adsorption of gas-phase species onto dust grains.

Thermal desorption is dependent on the binding energy, $E_i$, the dust temperature, $T_d$, and the vibrational frequency of a given species, $\nu_i$:

\begin{equation}\label{eq:desorp}
    k(E_i, \nu_i, T_d) = \nu_i \exp{\big(\frac{-E_i}{T_d} \big)}.
\end{equation}

\noindent The vibrational frequency is defined by:

\begin{equation}\label{eq:vib_freq}
    \nu_i = \sqrt{\frac{2\rho_S E_i}{\pi^2 m_i}},
\end{equation}

\noindent where $m_i$ is the specific mass of a given species and $\rho_S$ is a standard surface site density. Following \citet{Du..2014}, we assume a value of $\rho_S =10^{15}$ cm$^{-2}$. \reply{We adopt the binding energies of \citet{Garrod2008}; we note that our models are run at 400~K to ensure that ices have sublimated and the chemistry is dominated by gas phase reactions. }

For a given species $i$, the adsorption rate per unit volume per unit time is:

\begin{equation}\label{eq:adsorption}
    n_{\mathrm{dust}}n_i \sigma v_{t,i} s,
\end{equation}

\noindent where $n_{\mathrm{dust}}$ is the dust grains density,  $n_i$ is the density of the $i$ species, $\sigma$ is the average dust cross section, $v_{t,i}$ is the thermal velocity of species $i$, and $s$ is a sticking coefficient, which is calculated using the relationship described by \citet{Chaabouni..2012}. The number of dust grains are also used to keep the electric balance in the chemical network, i.e., we allow them to adopt one electric charge and recombine with electrons and ions.

\begin{table}[ht!]
\begin{center}
\caption{Different Physical Conditions Tested}
\label{tab:table2}
\begin{tabular}{ccc}
\hline
\hline
Designation & $\zeta$ (s$^{-1}$) & UV Field (G$_o$)\\
\hline
$\zeta_{12}$ & $10^{-12}$  & -\\
 $\zeta_{14}$ & $10^{-14}$  & -\\
$\zeta_{15}$  & \textbf{10$^{-15}$} & -\\
 $\zeta_{17}$  & 5$\cdot 10^{-17}$ &  - \\
 UV$_1$ & -  & 45\\
 UV$_2$ & - & 3.7\\
UV$_3$  & -  & \textbf{0.3}\\
 UV$_4$  & -  & 0.007 \\
\end{tabular}
\end{center}
\tablecomments{\reply{Fiducial models are marked in bold.}}
\end{table}

\begin{deluxetable*}{c|ccc|c}[ht!]
\tabletypesize{\scriptsize}
\tablewidth{0pt}
\tablecaption{Differences in \reply{Initial} Chemical Conditions Between Models \label{tab:deluxe}}
\label{tab:table3}
\tablehead{
\colhead{Model \#} & \colhead{CH$_4$ (C)}  & \colhead{CO} & \colhead{H$_2$O} & \colhead{C/O} }
\setlength\extrarowheight{3pt}
\startdata
{1} & ... & ...  & ... & 0.4  \\ 
{2} & 1.5$\cdot 10^{-4}$  & 10$^{-14}$ & 1.5$\cdot 10^{-4}$ & 1  \\ 
{3} & 1.5$\cdot 10^{-4}$  & 10$^{-14}$ & 7.5$\cdot 10^{-5}$ & 2  \\ 
{4} & 1.5$\cdot 10^{-4}$  & 10$^{-14}$ & 2.5$\cdot 10^{-5}$ & 6  \\
{5} & 1.5$\cdot 10^{-4}$ & 10$^{-14}$ & 1.5$\cdot 10^{-6}$ &100  \\ 
{6} & ...  & ... & 10$^{-8}$ & 1  \\ 
\enddata
\tablecomments{Boxes labeled by '...' mark that the species is simply our solar abundance \reply{specified in \ref{tab:table1}. All values listed are relative abundance. We tested the resulting chemistry using methane or atomic carbon as the extra carbon carrier in high C/O ratio scenarios}.}
\end{deluxetable*}

\subsection{\reply{Physical} Model Parameters} \label{subsec:parameters}

Table~\ref{tab:table1} provides the baseline set of initial chemical abundances used in our simulation based on typical cosmic abundances with a C/O ratio of 0.4 \citep{Nieva..2012}. 

The single-cell model also takes a variety of other physical conditions that we do not vary as input parameters. Our goal is to simulate the conditions that would exist on the X-ray and cosmic-ray dominated layers of disk surfaces in the inner few au near the star where the dust and gas temperatures exceed 400~K. 
\reply{A temperature of 400~K is chosen for our simulations} to match the temperature where gas phase water reactions would dominate the chemistry in these regions \citep{Kaufman1996, Bergin98}. 
We would then explore the reflexive response of carbon chemistry in this potentially water-rich environment. 

Additional conditions include a \reply{UV field},  extinction (A$_V$), number density, ionization rate, dust-to-gas ratio, and average grain size. For all of our models we set a grain radius of $10^{-2}$ cm, \reply{and a number density of 10$^{8}$ cm$^{-3}$}. Sub-mm dust grains usually carry most of the dust density is disk as shown by dust growth and evolution models \citep[e.g.][]{Birnstiel..2016}. \reply{These conditions (density and gas/dust temperature) are appropriate for the disk surface, which is detected by JWST \citep{Meijerink2009, Bosman2022}.} The time evolution of the simulation is run until 3 Myr with logarithmic steps assuming a ratio of 1.1.

\reply{We run two sets of models considering either ionization (X-ray)-driven chemistry and UV-driven chemistry. These are treated separately to resolve the key effects but also to understand the effects when one or the other dominates which will occur in different surface layers. For the X-ray ionization-driven runs we set the UV-driven photochemical reactions to zero.} \reply{Since we intend to replicate the surface conditions in UV-driven chemistry we set the dust-to-gas-ratio to $10^{-3}$ for the UV-driven chemistry models and $10^{-2}$ for ionization-driven ones. For the UV-driven models, we explore four different effective UV fields which are listed on Table \ref{tab:table2},  where our UV Field = 0.3 $G_o$ (UV$_3$) models serve as our fiducial case for subsequent analysis in \S~\ref{sec:result}. These were chosen based on where the photochemical reactions started to play a relevant role in gas-phase chemistry, and also at different locations where the infrared emission of some key species originates, as shown in the 2D radiation fields in \citet{Duval22}. }

\reply{In sum, our models are set up to explore the different cases of X-ray and UV dominance. Clearly this is a simplification. For instance, X-rays will be present in UV-dominated gas. In contrast UV-photons could be completely attenuated in X-ray dominated gas. For instance the attenuation depth of UV photons is $\sim10^{-3}$ g cm$^{-2}$ while X-rays can penetrate to 0.008 (1 keV) -- 1.6 (10 keV) g cm$^{-2}$ \citep{Bergin2007_ppv, Glassgold1997, Nomura2007}.
Regardless, our simulations will elucidate central aspects of the chemistry in both cases.}


We utilize cosmic-ray ionization as a proxy for X-ray ionization on disk surfaces, the reason for which is two-fold. First, to the first order, the effects of X-ray ionization of H$_2$ is comparable to cosmic-rays in that both produce H$_3^+$ \citep[e.g.,][]{Aikawa1999}. Second, observations of molecular ions suggest that cosmic-ray penetration is impeded in protoplanetary disks \citep{Cleeves2015, Aikawa2021, Seifert2021}. 
We survey cosmic-ray ionization rates, $\zeta$, over the range of $10^{-12}$, $10^{-14}$, $10^{-15}$, and $5\times10^{-17}$ s$^{-1}$, \reply{where our $\zeta = 10^{-15} s^{-1}$ ($\zeta_{15}$) model similarly serves as our fiducial case for comparison.}

To provide relevance to these ionization rates, we look at \citet{Glassgold1997} who model the X-ray ionization rate of a disk as a function of the stellar X-ray luminosity and \citet{Bethell2011} who provide parameterizations of the X-ray absorption cross-section. The vertical location of molecular emission from species such as C$_2$H$_2$ \citep[a key tracer of excess carbon;][]{Anderson21} and water is uncertain. Here we will be guided by thermal-chemical models optimized via analysis of the Spitzer emission of key molecules that emit from the inner few au where carbon grain destruction might occur. This analysis by  \citet{Bosman2022} suggests that molecular emission forms within gas with a vertical molecular hydrogen column density of $\sim$$5-30\times 10^{22}$ cm$^{2}$. 
\citet{Bethell2011} show that at 1 KeV, the X-ray cross section is of order of magnitude $10^{-22}$ cm$^{-2}$ assuming most of dust is settled in mid-plane and that the main absorbers are gaseous N$_2$ and CO. Thus, molecular emission arises at roughly the $\tau$ (1~keV) $\sim$ 5-30 range.   
In the models of \citet{Glassgold1997}, the X-ray ionization rate at $\tau$ (1~keV) $\sim$ 10 is of order $10^{-13}$ to $10^{-14}$ s$^{-1}$ depending on the plasma temperature. We thus explore ionization ranges that bound this range and extend to lower levels consistent with ionization dominated by cosmic-rays alone ($\zeta_{CR}$ $= 5 \times 10^{-17}$~s$^{-1}$). \reply{We note that this analysis is done using X-ray absorption cross-sections that separate the gas from the dust \citep{Bethell2011} and, in this instance, include only the gas. Thus the justification above for our ionization rate effectively incorporates significant dust evolution and settling to the midplane as expected for a Class II disk.}

We also set a constant temperature for the entirety of the simulation instead of calculating the thermal balance throughout the evolution of the disk.
Since we exploit a large extinction for our simulations, we do not include photodesorption reactions in the network as we expect them to be less relevant as the UV flux in the shielded regions also becomes negligible \reply{and in the UV-driven cases, the temperature is high enough to impede the adsorption of volatiles onto dust grains}. 

\subsection{Assumptions and Chemical Setup} \label{subsec:setup}

Table~\ref{tab:table3} shows the initial chemical conditions used in each model, with our solar model designated as Model 1.
Aside from CH$_4$, H$_2$O, and CO, the rest of the initial abundances as described in Table~\ref{tab:table1} are consistent for every successive model.
\reply{We choose the initial chemical conditions to represent different C/O ratios in the disks tracing different carbon enrichment. We do this by varying the abundances of water, carbon monoxide, methane, and/or atomic carbon. Among the different carbon enrichment scenarios, we test the possible differences in the chemistry for the cases where the carbon donor is both neutral carbon and methane. We repeat a C/O of unity in Models 2 and 6 to analyze the role that water has on the chemistry as an initial oxygen carrier. We note that a higher initial abundance of sulfur was also tested within the parameters of our models, but no significant effect on the overall chemistry was found.}

\subsection{Determining the Reaction Pathways}\label{subsec:reactions}
For each two-body gas-phase reaction in the network, the reaction rate scales with the respective abundances of the reactants.
Given the time dependence of the abundances, the most pertinent reactions for a specified species occur at the local maxima of the relative abundance of that species.
The dominant production and destruction reactions are given by the reactions with the largest corresponding reaction rate that has the species of interest as a product, and reactant, respectively.

This is determined by evaluating and exploring the highest terms of the gradient of these rates with respect to time.
The gradient is given by the Jacobian of the full chemical network, which when multiplied by the abundances of the reactants at a specific time gives the primary production reaction for the resulting product of that reaction. The negative of the gradient can be used in a similar manner to determine the main destruction reactions.
The species involved in the reactions can then be subsequently traced backward through their own most prevalent production reactions to ascertain the dominant reaction pathways.
While the physical conditions we choose result in models that straddle the line between \reply{negatively charged grains} or free electrons dominating, neither particle explicitly shows up in most of the principal pathways.

\section{Results} \label{sec:result}

\begin{figure*}[ht!]
\centering
\includegraphics[width=2.115\columnwidth]{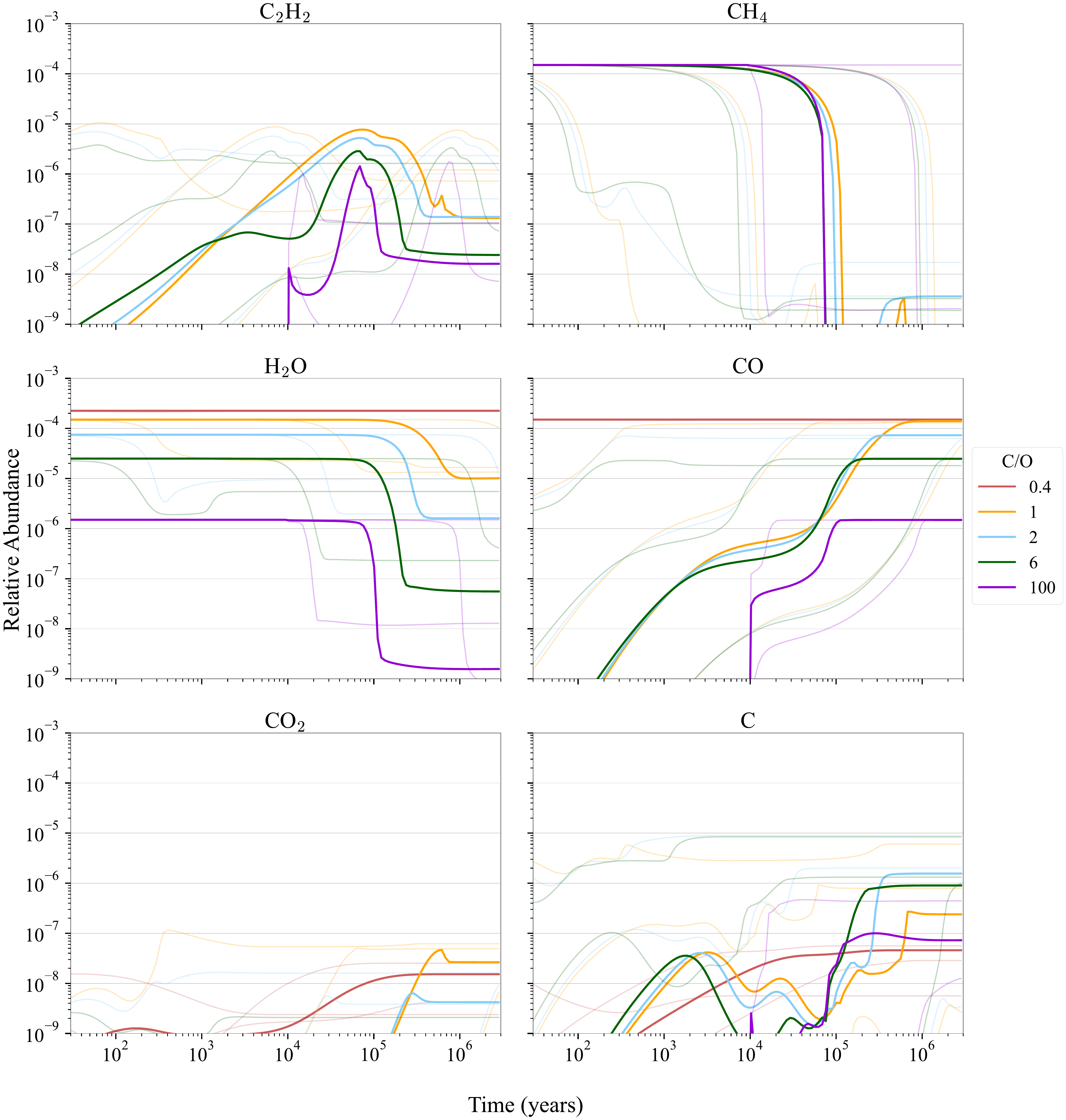}
\caption{Omnibus plots of C$_2$H$_2$, CH$_4$, \reply{H$_2$O, CO, CO$_2$, and C} showing the relative abundance of these species at every chemical condition for a number density of $10^{8}$ cm$^{-3}$ \reply{in the X-ray dominated case, sorted by initial C/O ratio. The solid lines represent the abundances of our simulations using our fiducial ionization rate $\zeta_{15}$ ($\zeta = 10^{-15} s^{-1}$), while the faded lines portray our other modeled rates ($\zeta_{12}$, $\zeta_{14}$, and $\zeta_{17}$). We find that the time at which the abundance of species such as C$_2$H$_2$ and CO reach their peak/equilibrium abundance is dependent upon the ionization rate, in that the higher ionization rates are represented by earlier peaks in the abundances of these molecules. The corresponding inverse effect is seen in CH$_4$ and H$_2$O, in which the lower ionization rate leads to a drop in its abundance later in the disk's lifetime or in some cases not at all.}}
\label{fig:ob_X-ray}
\end{figure*}

\begin{figure*}[ht!]
\centering
\includegraphics[width=2.115\columnwidth]{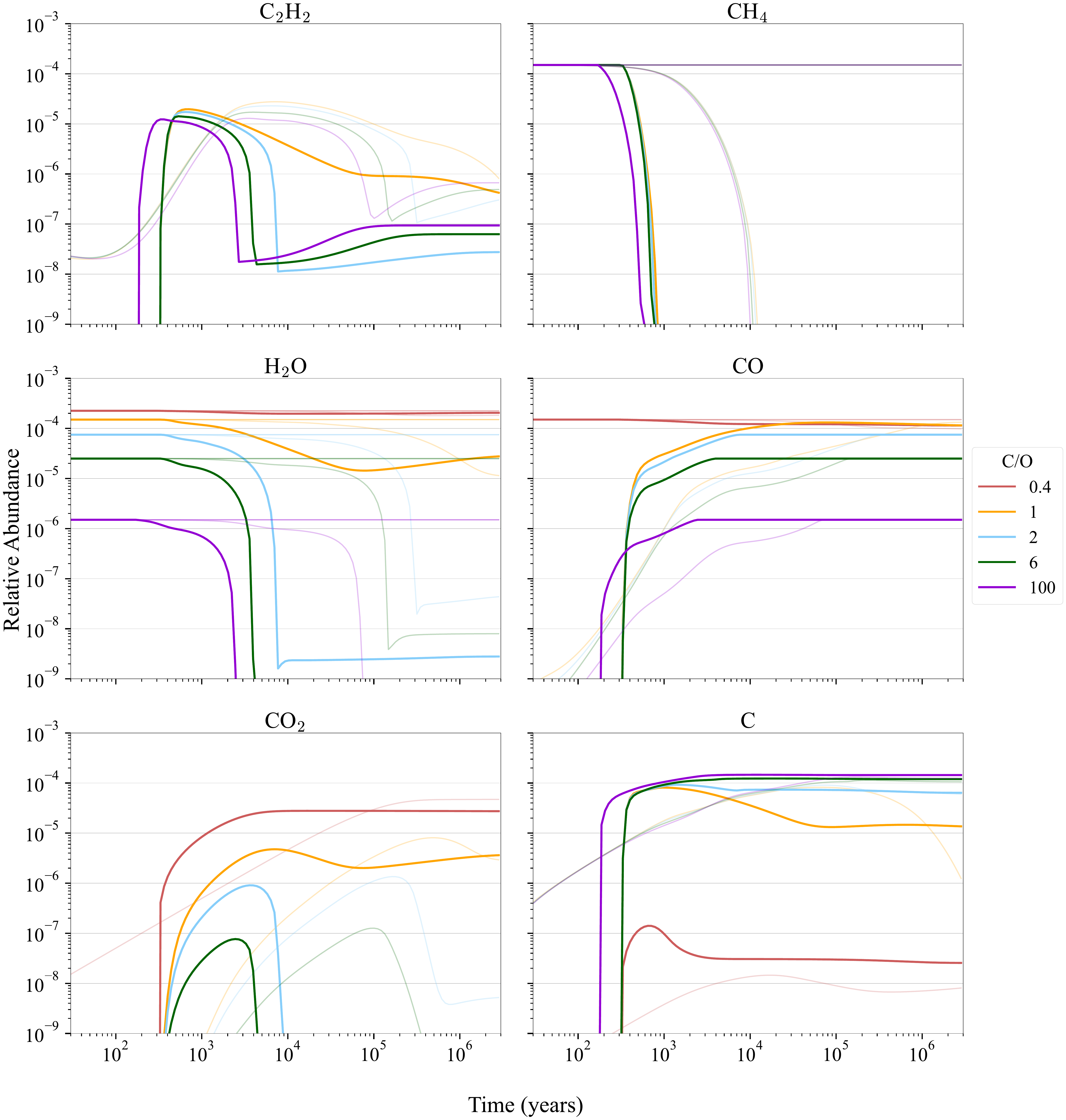}
\caption{\reply{Omnibus plots of C$_2$H$_2$, CH$_4$, \reply{H$_2$O, CO, CO$_2$, and C} showing the relative abundance of these species at every chemical condition for a number density of $10^{8}$ cm$^{-3}$ in the UV dominated case, sorted by initial C/O ratio, done in the same manner as Fig.~\ref{fig:ob_X-ray}, with our fiducial UV field instead being model UV$_3$ (UV field = $0.3$ $G_o$). The same effect described in Fig.~\ref{fig:ob_X-ray} can be seen here, where for example the peaks in C$_2$H$_2$ abundance in the non-fiducial models later in the disk's lifetime represents a smaller UV field. We note that for all non-CH$_4$ panels, only two physical models appear on the plot (UV$_3$ and UV$_4$); the reason for this is described in Fig.~\ref{fig:hm}.}}
\label{fig:ob_UV}
\end{figure*}

\subsection{Layout of Results} \label{subsec:layout}
Given their importance as building blocks for more complex molecules that also have emission lines at near-infrared and mid-infrared wavelengths, we put a particular focus on the following species: \reply{C$_2$H$_2$, CH$_4$, and CO$_2$}.

\reply{We find that these species respond distinctly to different stimuli with intricate dependencies, both chemically and physically. These complex dependencies are illustrated in Figs.~\ref{fig:ob_X-ray} and \ref{fig:ob_UV}, which show the relative abundance of these key species at our tested C/O ratios as a function of time. 
We label these figures as ``omnibus'' plots as they show, for the most part, all the model results for different ionization rates and C/O in one place.  As these plots may be confusing, we highlight the fiducial model ($\zeta_{15}$ and UV$_3$) for each ionization driver at a particular rate and display the rest of our models as faded lines in the background. In most cases it is clear that models with lower ionization rates/UV fields have chemical transitions to equilibrium at relatively later times, and vice versa. Outside of the cases where the strength of the ionization rate/UV field causes no virtually no chemistry to occur in the model (See Fig.~\ref{fig:hm}), the effect on the time of the chemical transitions is the primary effect of this parameter. We find that the impact of the different ionization rates/UV fields on molecular equilibrium abundances is less than that of our different C/O ratios, with the only noticeable trend being an increase in equilibrium abundance for earlier chemical transitions (strong ionization rates/UV fields).}
Below, we discuss the causes of these plots for C$_2$H$_2$, CH$_4$, and CO$_2$, as well as a few other hydrocarbons in \S~\ref{subsubsec:other}.

\subsubsection{C$_2$H$_2$} \label{subsubsec:c2h2}


\reply{C$_2$H$_2$ is a prime species for study in particular as we find that most models push a modest (few percent) fraction of the excess carbon into C$_2$H$_2$. \reply{Our most important conclusion, as discussed further below in \S~\ref{subsec:water implications}}, is that C$_2$H$_2$ shows a preference for the presence of H$_2$O.

\reply{In our models C$_2$H$_2$ is not formed at high abundance levels when C/O = 0.4 (Model \#1).  This differs from \citet{Duval22} who find that C$_2$H$_2$ can form in these conditions.  In our case this is because we do not include CO photodissociation which would generate carbon to form C$_2$H$_2$ while the oxygen is rapidly placed into water as noted by \citet{Duval22}.

Beyond our C/O = 0.4 model (Model \#1), C$_2$H$_2$ is optimally produced at a C/O = 1 (Model \#2), with a lower equilibrium abundance as the C/O increases. This effect occurs in both our X-ray and UV models. The reason for this is that while C$_2$H$_2$ acts as an initial carbon bearer (peaking at approximately $10^5$ years in the X-ray regime and $10^3$ years in the UV regime), it also serves as a donor to the carbon chemistry in the later years of the disk's lifetime.}

The leading production reaction of this species in our X-ray models is
\begin{equation} \label{eq:1}
\mathrm{H_2O  +  C_2H_3^+  \rightarrow  C_2H_2  +  H_3O^+,}
\end{equation}
exemplifying the C$_2$H$_2$ dependence upon water covered in \S~\ref{subsec:water implications}. This explains why, as seen in Figs.~\ref{fig:ob_X-ray} and \ref{fig:ob_UV}, C$_2$H$_2$ is preferentially produced in our C/O = 1 and 2 models (Models \#2 and 3), as these are the models with the highest initial amount of H$_2$O put into the system.
We find that the subsequent dominant destruction reaction that distributes the C$_2$H$_2$ into these species is
\begin{equation} \label{eq:2}
\mathrm{H_3^+  +  C_2H_2  \rightarrow  C_2H_3^+  +  H_2.}
\end{equation}
In the X-ray regime, this effect of C$_2$H$_2$ as a carbon donor can also be seen in in Fig.~\ref{fig:longchain}, in which this carbon goes into the longer chain hydrocarbons such as C$_5$, C$_6$, C$_6$H, C$_8$, and C$_8$H.

In the UV regime, the primary production reaction of C$_2$H$_2$ in the early years of the disk follows as 
\begin{equation} \label{eq:3}
\mathrm{H_2  +  C_2H  \rightarrow  C_2H_2  +  H,}
\end{equation}
with a subsequent primary destruction reaction of
\begin{equation} \label{eq:4}
\mathrm{C_2H_2  +  C_3H^+  \rightarrow  C_5H_2^+  +  H.}
\end{equation}
\reply{This formation path is also highlighted by \citet{Kanwar2024}, which illustrates the importance of H$_2$ in our UV models.}
Instead of fostering the formation of long chain hydrocarbons as seen for the X-ray dominated case, in this regime the carbon is instead predominantly forced into C and CO.
With the exception of our C/O = 0.4 model (Model \#1), C$_2$H$_2$ peaks at a relative abundance above $10^{-6}$ in the majority of our X-ray regime, and above $10^{-5}$ in our UV regime early in the lifetime of the disk. C$_2$H$_2$ similarly settles at an equilibrium abundance anywhere from $10^{-8}$ to $10^{-6}$ in all of these models, implying it contains up to approximately 2\% of the total carbon reservoir, as further discussed in \S~\ref{subsec:reservoirs}.

\subsubsection{CH$_4$} \label{subsubsec:ch4}
CH$_4$ \reply{is a} transient species in that when present at high abundances ($10^{-4}$), \reply{it} will eventually tend to fall several orders of magnitude as \reply{its} destruction reactions become dominant (see Figs.~\ref{fig:ob_X-ray} and \ref{fig:ob_UV}). \reply{Its} lifespan in the gas phase is directly dependent on the ionization rate \reply{and UV field}, i.e., in highly ionized environments \reply{it is} very short-lived, while \reply{it is} more prevalent for low ionization conditions.
It follows that observability of an extreme transient such as CH$_4$ in both regimes preferentially occurs before the prime cascade of chemistry, as it is a donor molecule that can help drive the formation of other species. The dominant destruction reaction of CH$_4$ that adds its carbon to the chemistry in the X-ray case is
\begin{equation} \label{eq:6}
\mathrm{H_3^+  +  CH_4  \rightarrow  CH_5^+  +  H_2,}
\end{equation}
while for our UV regime is
\begin{equation} \label{eq:6-1}
\mathrm{H  +  CH_4  \rightarrow  CH_3  +  H_2.}
\end{equation}

In Figs.~\ref{fig:ob_X-ray} and \ref{fig:ob_UV}, we can observe the consequences of this phenomenon in the ubiquitous presence of CH$_4$ at an abundance of 10$^{-4}$ in a handful of our non-fiducial ionization rate models. These faded abundance lines correspond to our low ionization models ($\zeta_{17}$), leading to the consistent presence of CH$_4$ throughout the lifetime of the disk due to the relatively longer ionization timescales. We note that in these models, without the donation of the carbon from CH$_4$ to the chemistry, little-to-no chemistry occurs in these very specific models and all of our non-initially abundant carbon species virtually do not get produced.

In the same manner that C$_2$H$_2$ is optimally produced at a C/O = 1 (Model \#2), we find that the equilibrium abundance of CH$_4$ is highest at a C/O = 2 (Model \#3). The production reaction that reproduces CH$_4$ after the initial kick into the chemistry follows as
\begin{equation} \label{eq:7}
\mathrm{H_2  +  CH_3  \rightarrow  CH_4  +  H}
\end{equation}
in the X-ray regime, once again underscoring the importance of H$_2$ in our models.

\subsubsection{CO$_2$} \label{subsubsec:co2}
CO$_2$ is found at significant abundances (greater than $10^{-8}$) in all of our \reply{UV} models, while hardly being produced in our X-ray models. We find that in our UV models, CO$_2$ is predominantly produced through the reaction
\begin{equation} \label{eq:8}
\mathrm{OH  +  CO  \rightarrow  CO_2  +  H}.
\end{equation} 
\reply{Reaction~\ref{eq:8} is also the dominant reaction in the majority of our X-ray models, but is simply at a much lower reaction rate due to the different chemistry. We can also see from Fig.~ \ref{fig:ob_UV}, that as the C/O decreases, the abundance of CO$_2$ increases, being optimally produced at a C/O = 0.4 (Model \#1) to an equilibrium abundance on the order of $\sim$10$^{-5}$. This makes sense, as in Reaction~\ref{eq:8}, CO directly aids in this boost in production, the initial abundance of which is amplified in our low C/O models.

In our higher C/O ($>$ 1) models, and as can be noticeably seen in Fig.~\ref{fig:ob_UV}, the CO$_2$ markedly decreases after its peak abundance around $5 \times 10^{3}$ years.
This is due to the primary destruction reaction
\begin{equation} \label{eq:10}
\mathrm{H^+  +  CO_2  \rightarrow  HCO^+  +  O}.
\end{equation} 
As with Reaction~\ref{eq:8}, this is the same reaction responsible for the destruction of CO$_2$ in the X-ray scenario (around $5 \times 10^{3}$ years), except to a much smaller extent.}

\subsubsection{Atomic Carbon \& Other Hydrocarbons} \label{subsubsec:other}
\subsubsection{C} \label{subsubsubsec:c}
\reply{We find that C is highly produced in our UV-dominated scenario, being produced at a peak abundance on the order of 10$^{-4}$. This shows, as also seen in the second panel of Fig.~\ref{fig:longchain}, that C, as well as CO, serve as the carbon sinks in our UV regime. C is optimally produced in our higher C/O models, largely through the production reaction
\begin{equation} \label{eq:11}
\mathrm{H + CH \rightarrow C + H_2.}
\end{equation}
As can be seen in our C/O = 1 model (Model \#2) in the UV case (Fig.~\ref{fig:ob_UV}), the equilibrium abundance is approximately one order of magnitude lower than its peak abundance, which we find is realized by the destruction reaction
\begin{equation} \label{eq:12}
\mathrm{C + H_3O^+ \rightarrow HCO^+ + H_2}.
\end{equation}}
\reply{We note that these models do not create ionized carbon as our single point solution was placed optimally to allow for molecular formation.  In this layer the higher densities fosters electron recombination which limits the formation of C$^+$.}

In our X-ray scenario, we find that atomic carbon is optimally produced between a C/O = 2 and 6 (Models \#3 and 4), produced to a relative abundance of only approximately 10$^{-6}$, highlighting the fact that C is preferentially produced in the UV regime.

\subsubsection{CO} \label{subsubsubsec:co}
We find that CO is consistently a highly stable molecule in both our X-ray and UV models, showing a preference in both cases for a lower C/O (ignoring the C/O = 0.4 model, where CO is initially put into the system). This consistent presence is accomplished through the production reaction
\begin{equation} \label{eq:14}
\mathrm{H_2O + HCO^+ \rightarrow CO + H_3O^+}
\end{equation}
in both scenarios. Within our higher CO models where H$_2$O is not as present, the supplemental reaction
\begin{equation} \label{eq:14-1}
\mathrm{HCN  +  HCO^+  \rightarrow  HCNH^+  +  CO}
\end{equation}
also aids in the production of CO in the X-ray case, while the reaction
\begin{equation} \label{eq:14-2}
\mathrm{C + HCO^+ \rightarrow CO + CH^+}
\end{equation}
takes over in the analogous UV case. As also seen in Fig.~\ref{fig:longchain}, CO serves as the primary carbon sink in both the X-ray and UV-driven chemistry regimes.

While not visible through Figs.~\ref{fig:ob_X-ray} and \ref{fig:ob_UV}, CO also serves as a slight donor to the chemistry through the destruction reaction
\begin{equation} \label{eq:15}
\mathrm{H_3^+ + CO \rightarrow HCO^+ + H_2}
\end{equation}
in the X-ray scenario and
\begin{equation} \label{eq:15-1}
\mathrm{OH + CO \rightarrow CO_2 + H}
\end{equation}
in the UV regime, though this does not significantly affect the equilibrium abundance of CO. We note that as CO serves as such a stable carbon sink in our models, this results in our C/O = 0.4 model (Model \#1) exhibiting little to no carbon chemistry. This is in contrast to the rest of our models, in which virtually all of the carbon is put in CH$_4$, which as discussed in \S~\ref{subsubsec:ch4}, is an extremely transient donor species to the chemistry. This makes the lack of C/O = 0.4 lines present in Figs.~\ref{fig:ob_X-ray}, \ref{fig:ob_UV}, \ref{fig:hm}, \ref{fig:other_hcarb_xray}, and \ref{fig:other_hcarb_uv} a consequence of our choice in initial conditions rather than one inherent to the chemistry.

\subsubsection{C$_2$H$_4$} \label{subsubsubsec:c2h4}
As seen in both Figs.~\ref{fig:other_hcarb_xray} and \ref{fig:other_hcarb_uv}, C$_2$H$_4$ does not get produced to a detectable equilibrium abundance in either case, despite the fact that we have seen it in recent observations using JWST \citep{Arabhavi2024, Kanwar2024}. However, in the X-ray regime, our C$_2$H$_4$ abundance peaks at approximately 10$^{-5}$ in all of our intermediate C/O models (100 $>$ C/O $>$ 0.4).
This is due to the production reaction
\begin{equation} \label{eq:16}
\mathrm{H_2O + C_2H_5^+ \rightarrow C_2H_4 + H_3O^+}.
\end{equation}
The subsequent destruction reactions that prevent C$_2$H$_4$ from surviving the lifetime of the disk follow as
\begin{equation} \label{eq:16-1}
\mathrm{H_3^+ + C_2H_4 \rightarrow C_2H_3^+  +  H_2  +  H_2}
\end{equation}
and
\begin{equation} \label{eq:16-2}
\mathrm{H_3^+ + C_2H_4 \rightarrow C_2H_5^+  +  H_2}.
\end{equation}}



\section{Discussion} \label{sec:discussion}

\subsection{X-Ray vs. UV-Dominated Chemistry} \label{subsec:x-uv}

\begin{figure*}[ht!]
\centering
\includegraphics[width=1\textwidth]{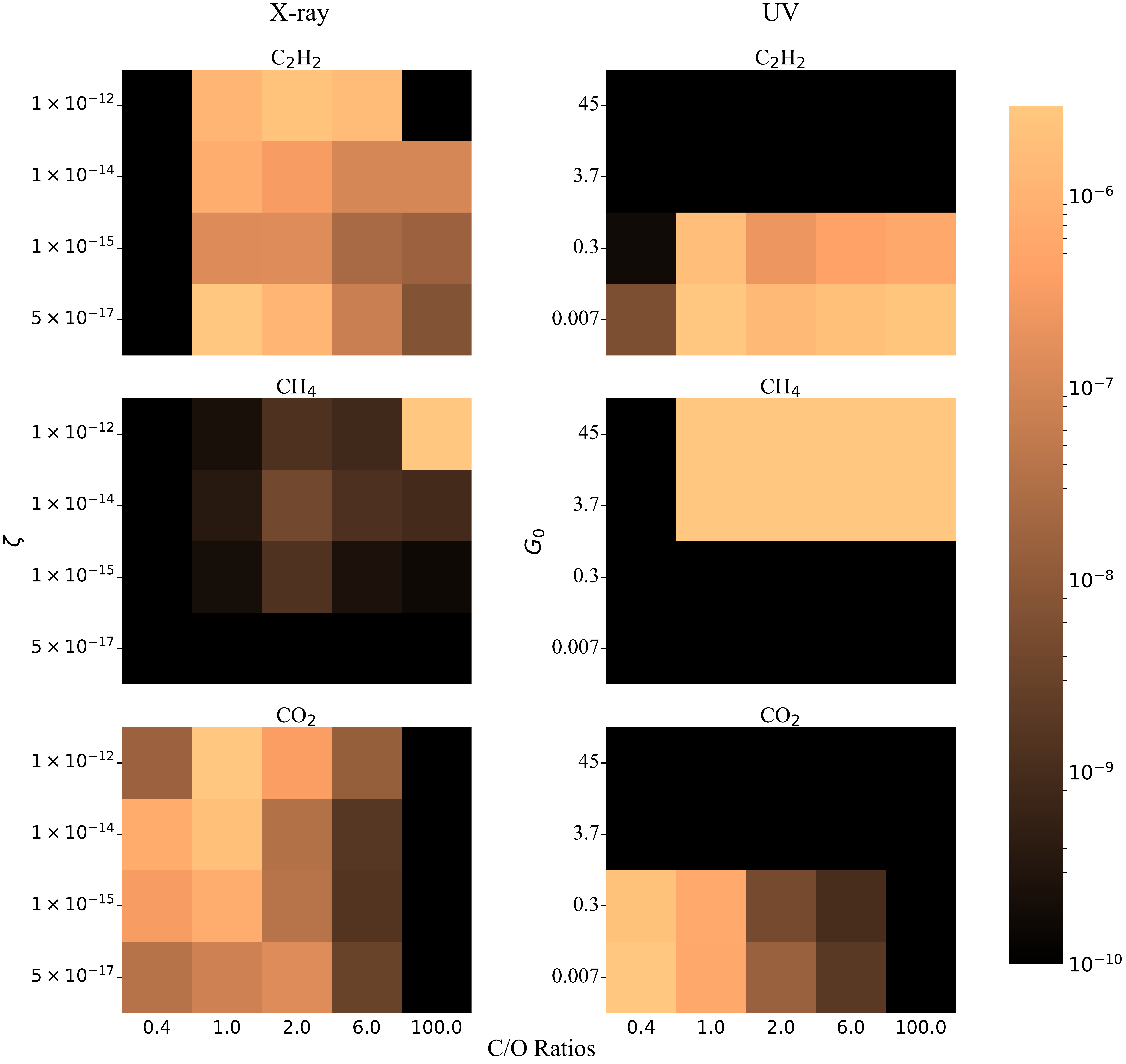}
\caption{\reply{Heat map of key hydrocarbons abundance at $3\times10^6$ years for different C/O ratios and physical conditions in both the X-ray and UV-dominated chemistry regimes. We find that in our high UV field models (UV$_1$ and UV$_2$), virtually no chemistry occurs whatsoever, as can also be seen by the lack of non-fiducial models in Fig.~\ref{fig:ob_UV}. This is because in our UV models, the production of most non-initially abundant species (e.g. C$_2$H$_2$, CO$_2$) is followed by a subsequent drop in abundance, most strongly effecting our higher C/O models. In our high UV field models, this drop would occur before the production of these species begins, effectively killing the chemistry before it can happen, resulting in the equilibrium abundances equal to the initial abundances in these models in the above heat map.}
\label{fig:hm}}
\end{figure*}

\begin{figure*}[ht!]
\centering
\includegraphics[width=2.115\columnwidth]{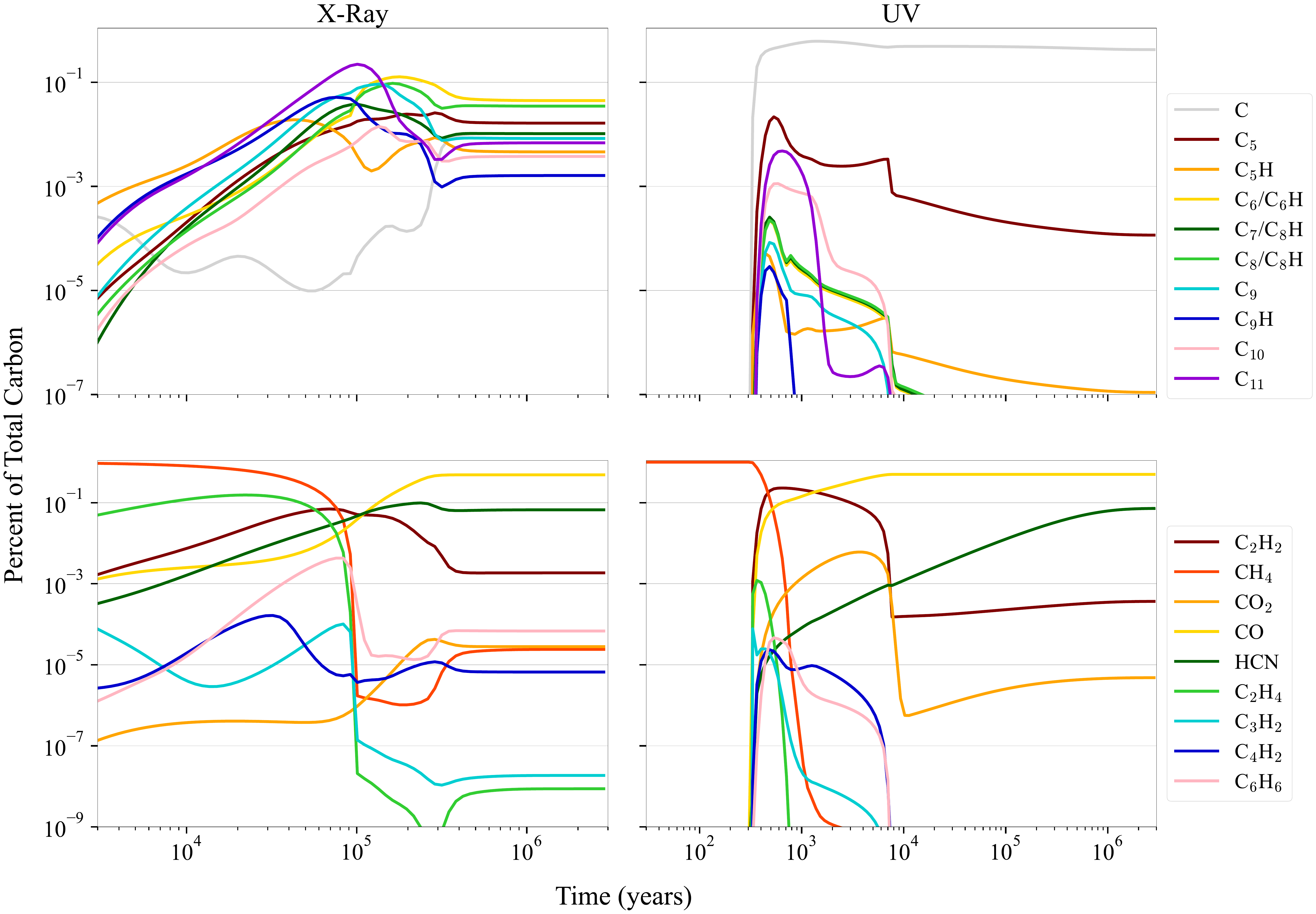}
\caption{\reply{4-panel individual species plot of the amount of carbon in each carbon-bearing species for both the UV and the X-ray case. All panels are displayed at our fiducial $\zeta_{15}$ ($\zeta = 10^{-15} s^{-1}$) ionization rate at a C/O = 2 (Model \#3). This plot changes as C/O increases (not shown) only in that the longer carbon chain molecules become even more dominant above CO in the X-ray regime, and that C takes over CO as the primary carbon sink in the UV regime.}}
\label{fig:longchain}
\end{figure*}

\reply{In our 1D models we have isolated a few essential aspects that delineate differences between a UV dominated and X-ray dominated models. Of course, true disks have both UV and X-ray photons present, and these solutions can provide some guidance for the key energetics that are activating the observed chemistry.
Fig.~\ref{fig:hm} shows the difference in the equilibrium abundances of key hydrocarbons between our X-ray and UV-driven models, demonstrating the effect different chemistry-drivers have on the observable chemical abundances.
Fig.~\ref{fig:longchain} shows the reason for these differences using our illustrative fiducial models ($\zeta_{15}$/$UV_{15}$ Model \#3) in terms of the abundances of the primary carbon reservoirs as a function of time for an identical elevated C/O ratio with a chemistry powered separately by X-rays and UV photons. These two models are representative of solutions that provide the most productive (i.e. carbon-rich) chemistry. In this figure we provide the full extent of the chemistry in the sense that we track where the excess carbon is deposited in equilibrium but also show the time evolution of key species that are observed by JWST to be present in the sources with hydrocarbon-rich spectra (e.g. C$_2$H$_2$, C$_4$H$_2$, C$_2$H$_4$, CH$_4$, and C$_2$H$_6$). These species are, at times, major carbon reservoirs but only on specific timescales that scale with the ionization rate ($\zeta$), or UV field. Smaller ionization rates mean longer timescales as seen in Fig.~\ref{fig:ob_X-ray} and Fig.~\ref{fig:ob_UV}.}

\reply{A few general statements can be made from the example in Fig.~\ref{fig:longchain}. (1) Both models take whatever oxygen is available to produce CO in equilibrium.    
(2) X-rays are generally more productive and produce a longer-lived hydrocarbon chemistry than UV photons. This may suggest that portions of the hydrocarbon emissions arise from columns where the chemistry is powered by X-rays.  This is consistent with a recent analysis by \citet{Woitke2024}. A key tracer of this result is CH$_4$ which is present for short timescales (typically $< 10^5$~yrs) for both X-ray and UV solutions but timescales are much shorter for the UV dominated solutions (compare CH$_4$ panels in omnibus plots in Fig.~\ref{fig:ob_X-ray} and Fig.~\ref{fig:ob_UV}). (3) Looking at the observable tracers (Fig.~\ref{fig:longchain}), C$_2$H$_2$ reaches higher abundance levels $>$10\% of the carbon for small periods of time in the UV dominated case. The high C$_2$H$_2$ columns seen in some low mass (M star) sources \citep{Tabone2023, Arabhavi2024, Kanwar2024} could arise from a UV-dominated chemistry. This equilibrium is short-lived unless the carbon supply terms, whether vertical mixing of carbon grain destruction from the midplane or radial advection of carbon-rich gas from the outer disk, are operative on similar, or shorter, timescales. These timescales are dependent on the ionization rate of this gas which can likely be constrained via forward models such as those presented by \citet{Woitke2024}.   (4) In these models the X-rays appear to be more productive for other detectable species such as C$_4$H$_2$, C$_2$H$_4$, and C$_2$H$_6$ (see Figs.~\ref{fig:other_hcarb_xray} and \ref{fig:other_hcarb_uv}. (5) A final point is that the X-ray solutions end with an equilibrium that places much of the carbon in long-chain species (e.g. C$_5$, C$_6$, C$_6$H, C$_8$, C$_8$H) as seen in other simulations \citep{Wei2019}. In contrast, the UV dominated solutions place significant carbon in atomic form. This may be a limitation of these single point models as other deeper layers will exist with less UV.}

\subsection{The C/O Ratio and Carbon Supply} \label{subsec:c_o}

\reply{In our models we have explored C/O ratios from interstellar gas to 100.  Based on existing analyses, the former may be typical in some systems \citep{Gasman2023}, modest C/O values are consistent with some hydrocarbon rich systems \citep{Kanwar2024, Arabhavi2024}, while the highest value is perhaps an extreme end of the solutions. A key point in these solutions is that the highest abundances of the observed hydrocarbons exhibits significant time dependencies for typical (and expected) conditions for both X-ray and UV powered chemistry. }

\reply{One surprising result is found for C$_2$H$_2$ which exhibits its highest abundance for both X-ray and UV when C/O = 1 (Model \#2). This is somewhat counter intuitive as a generic expectation would suggest that excess carbon in carbon-rich gas would likely lead to efficient C$_2$H$_2$ production.  However, the overall chemistry rapidly parses the carbon into carbon chains for X-ray or into C/C$_2$ for UV (Figs.~\ref{fig:other_hcarb_xray} and \ref{fig:other_hcarb_uv}). This dependence may be the result of the link between C$_2$H$_2$ formation and water which is discussed in \S~\ref{subsec:water implications}. Another interesting aspect of the C$_2$H$_2$ chemistry is that the time C$_2$H$_2$ spends near its peak abundance decreases from C/O = 1 (Model \#2: broad peak) to C/O = 100 (Model \#5: narrower peak) regardless of ionization source.}

\reply{Looking at the other observable hydrocarbons it is a general statement that X-ray chemistry is more favorable for their formation as can be seen by comparing Fig.~\ref{fig:other_hcarb_xray} to Fig.~\ref{fig:other_hcarb_uv}. However, abundant C$_2$H$_2$ is found over long timescales when the UV field is weaker (e.g. UV$_4$) as seen in Fig.~\ref{fig:ob_UV}. This longevity occurs for C$_2$H$_2$ but is not replicated for other long chain hydrocarbons in these models as they appear to be rapidly dissociated (Fig.~\ref{fig:other_hcarb_uv}). In general molecules such as C$_2$H$_4$, C$_4$H$_2$, and C$_6$H$_6$ appear to favor formation in X-ray dominated gas as seen in Fig.~\ref{fig:other_hcarb_xray} on timescales of $\sim 10^4 - 10^6$~yrs depending on the ionization rate. We note that the lowest ionization rate of $\zeta = 5 \times 10^{-17}$ ($\zeta_{17}$) provides the longest timescales and high abundances for C$_6$H$_6$ (as one example). This particular molecule (benzene) also preferentially forms in higher abundances when the C/O ratio is elevated.}


\subsection{Comparisons to Literature} \label{subsec:compare}

We can also compare our results to \citet{Wei2019}, who find that long chain carbon species such as C$_9$, C$_9$H$_2$, and C$_{10}$ come to be the most abundant at 1 au, we also find that these species do not reach detectable abundances (greater than $10^{-8}$) and especially come to dominate the chemistry in the enhanced carbon-carrier models (see Fig.~\ref{fig:longchain}. \citet{Wei2019} also find that the carbon in CH$_4$ is primarily transferred into the species C$_4$H, \reply{C$_5$H}, and C$_6$H, while we have found that C$_4$H does not appear in our models, even in the X-ray scenario. \reply{It is possible that these differences could be due to changes in the UMIST database from 2006 \citep{Woodall2007} as used by \citet{Wei2019} and our use of UMIST 2012 \citep{McElroy2013}.}

In the case of \citet{Kanwar2023}, they expanded the chemical reaction pathways to include three body reactions, including the formation of simple PAHs. This can lead to vastly different reaction pathways than the simple ones we explore in \S~\ref{sec:result}. \reply{However, this more detailed chemical network which includes larger hydrocarbons does not significantly affect the observed fluxes of C$_2$H$_2$, meaning that some other process must be responsible}.

\subsection{Transients and Reservoirs} \label{subsec:reservoirs}

In our models we find that CH$_4$ is readily destroyed and is effectively a transient carbon-donor species in this chemistry.
As discussed in \S~\ref{subsec:compare}, the carbon from CH$_4$ primarily goes into CO, C$_2$H$_2$, and HCN. Figs.~\ref{fig:ob_X-ray} and \ref{fig:ob_UV} shows that a transient species such as CH$_4$ is consistently destroyed, while CO, C$_2$H$_2$, and HCN indeed dominate in almost all of these conditions. More directly it is indicative of the fact that the detection of CH$_4$ within the context of X-ray-driven chemistry requires that this species be initially abundant or constantly supplied to the system, as its destruction timescales can be quite short. We note that different results may be found in models where CO photodissociation may ignite a constant CH$_4$ formation.
In this regard, in the limit of X-ray-driven chemistry, the detection of hydrocarbons such as C$_4$H$_2$, C$_2$H$_2$, and CH$_4$ in the absence of water \citep{Tabone2023} would imply that CH$_4$ must have been pre-supplied into the system \citep{Mah2023}. However, it is possible and indeed likely that UV is a major player in contributing to this chemistry as discussed in \citet{Kamp2023}.

In our models another species that stands out is HCN (shown in Appendix~\ref{appendix:a}), which becomes an important and stable reservoir of carbon. This is of interest as in the seminal paper by \citet{Pascucci2009}, they delineate a significant ``underabundance of HCN relative to C$_2$H$_2$'' in the atmospheres of \reply{disks around} cool stars compared to the more massive solar mass counterparts. Based on the Chandra Survey of Orion, for young stars there is no clear difference in X-ray luminosity \citep{Preibisch2005}. In our simulations, in the context of carbon release in relatively water-poor gas, we do not see appreciable changes in the HCN abundance. To match the Spitzer result from \citet{Pascucci2009} it may require chemical differences (e.g. N$_2$ vs. NH$_3$, higher water content, or lower carbon content), a combination of X-ray and UV-driven chemistry, or perhaps an excitation effect.

\subsection{Importance of Water in the Context of Infrared Observations} \label{subsec:water implications}
\begin{figure*}[ht!]
\centering
\includegraphics[width=1\textwidth]{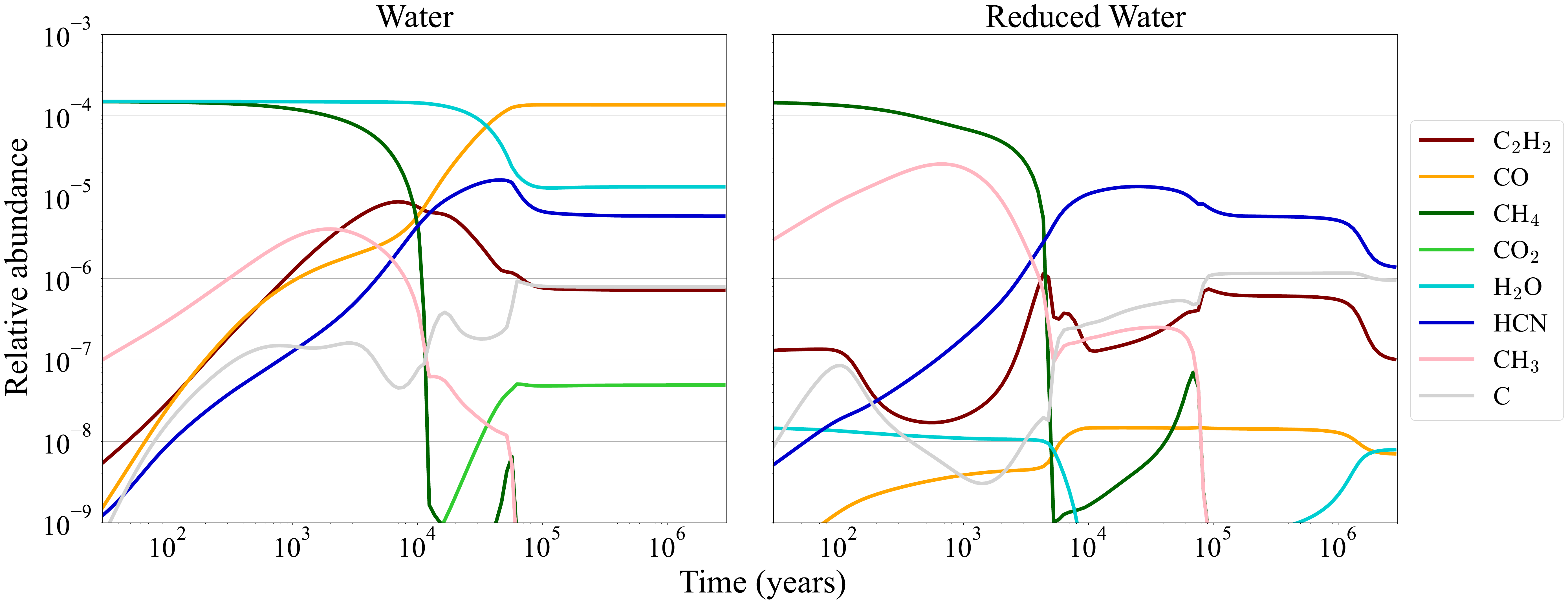}
\caption{\reply{Abundance over time plot showing the production of several hydrocarbons with and without the presence of water (Models \#2 vs. 6). This model is evaluated in the X-ray dominated case at our $\zeta_{14}$ ($\zeta = 10^{-14} s^{-1}$) ionization rate.}
\label{fig:water}}
\end{figure*}

Fig.~\ref{fig:water} demonstrates the role that the initial amount of water-ice has on individual species. In general, the presence of oxygen within the chemical equilibrium results in the destruction of hydrocarbons and the placing of oxygen into CO in the X-ray-driven case (see Fig.~\ref{fig:ob_X-ray}). However, this is not the case in our simulations when oxygen is in the form of water. This might be in part because the high temperature fuels the fast gas phase formation of water \citep{Wagner1987, Kaufman1996}, leaving little free atomic oxygen in the gas phase. This was recently shown to be a key fact in allowing for a rich chemistry to exist in the inner few au of a thermochemical model constraining an average Spitzer spectrum of a T Tauri disk \citep{Duval22}.

In our work, we find that when the system is in the presence of H$_2$O, the oxygen primarily goes into CO, just as the carbon goes into CO, C$_2$H$_2$, and HCN as discussed in \S~\ref{subsec:compare}. This effect can be seen in Fig.~\ref{fig:ob_X-ray}, in which the solar levels of initial water (10$^{-4}$) clearly enhance the equilibrium abundance of CO. This speaks to the stability of CO at approximately 1 au, as well as the dominance of C$_2$H$_2$ and HCN from \S~\ref{subsec:reservoirs}. We find that C$_2$H$_2$ in particular thrives in a high-water environment through the neutralizer Reaction~\ref{eq:1}.
H$_2$O does not primarily serve as a hydrogen donor to the chemistry, but rather an oxygen donor to chemical reactions such as Reactions~\ref{eq:1}. \citet[Table~5]{Najita2011} find a much larger C$_2$H$_2$ column \reply{density} than we find in our models, demonstrating that C$_2$H$_2$ is even more stable in their models. 

Fig.~\ref{fig:water} shows that the presence of water leads to abundant levels of C$_2$H$_2$, which is consistent with the mid-infrared observations of isolated systems showing both water and  C$_2$H$_2$ features \citep[e.g][]{Grant..2023, Tabone2023, Gasman2023}. On the contrary, in cases when water is not abundant or depleted, we observe a slight increase in the abundance of CH$_3$, CH$_4$ (see Fig.~\ref{fig:water}, and CH$_3^+$ (not shown), an effect that is enhanced if the chemistry of vibrationally excited H$_2$ is included \citep{Agundez..2010}. CH$_3^+$ has only been reported to be present in a water-poor system \citep{Berne2023},  in particular, embedded in a massive star-forming environment with severe UV irradiation such as the Orion Bar. This trend possibly points to a distinct chemistry between typically isolated systems and the most typical star formation environments being heavily irradiated by massive stars.

\subsection{Methane vs. Atomic Carbon as an initial carrier} \label{subsec:carrier}

\begin{figure*}[ht!]
\centering
\includegraphics[width=1\textwidth]{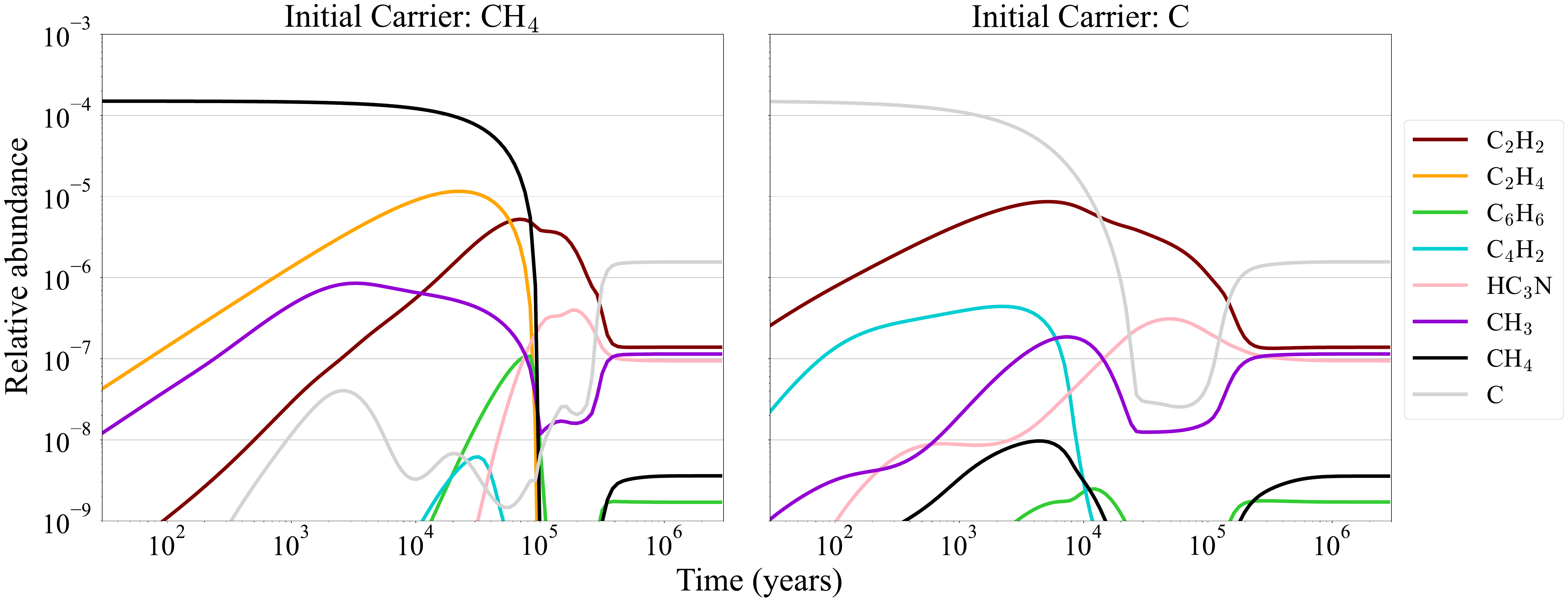}
\caption{\reply{Abundance over time plot showing the production of several hydrocarbons with CH$_4$ vs. C as an initial carrier. This model is evaluated in the X-ray dominated case at our fiducial $\zeta_{15}$ ($\zeta = 10^{-15} s^{-1}$) ionization rate at a C/O = 2 (Model \#3).}
\label{fig:carrier}}
\end{figure*}

\reply{In our models we have explored solutions where methane is effectively the carrier of excess carbon. This assumption may be compatible with some solutions for the elevated C/O ratio \citep[e.g.,][]{Mah2023}, but it may not be consistent with the destruction of refractory material which could follow a different path and lead to a range of products. From the observations one aspect is abundantly clear: C$_2$H$_2$ has the highest observed column density and is a prime beneficiary of this chemistry, regardless of its origin \citep{Tabone2023, Arabhavi2024, Kanwar2024}.  } 

\reply{To explore one potential alternative we proved one simulation in the X-ray dominated case ($\zeta_{15}$ Model \#3) where we place the excess carbon in atomic carbon. This is shown in Fig.~\ref{fig:carrier} where we compare models where the initial carrier of excess carbon is CH$_4$ versus C. There are modest differences in the composition as a function of time. If the initial excess carbon carrier is methane, then more complex hydrocarbons are present in abundance earlier in the evolution (e.g., C$_2$H$_4$).  However, the central point is that both models attain the same equilibrium composition on similar timescales ($\sim$10$^5$~yrs in this instance); this timescale is set by the overall ionization rate and would be shorter for higher levels of ionization.}

\section{Conclusion} \label{sec:conclusion}
In this work, we have tested the implications of \reply{different C/O ratios and chemistry drivers} in disks considering conditions inside terrestrial planet-forming regions traceable by infrared wavelengths. By flooding the system with \reply{carbon} through an assortment of individual \reply{C/O ratios} and physical conditions, we analyze how these distinct initial conditions affect the overall chemistry, powered \reply{by cosmic/X-rays and/or a UV field,} through the relative abundance of key molecular tracers and lead to the findings of current ongoing observational campaigns of inner disk chemistry with the JWST. 

\reply{We have found that production of hydrocarbons is extremely dependent upon the chemistry driver of our models, with species such as CO, C$_2$H$_2$, and HCN along with long chain hydrocarbons such as C$_5$, C$_6$, and C$_8$ being preferentially produced in the X-ray dominated scenario. Similarly, CO, C, HCN, and CO$_2$ are predominantly produced in the UV-driven scenario, with the former two acting as sinks for the vast majority of the carbon put into the system.}

We have \reply{also} found that the initial water abundance plays a key role in the abundance and chemical paths of several hydrocarbons such as \reply{C$_2$H$_2$ and C$_4$H$_2$, where the former can hold up to approximately 2\% of the total carbon supply.} We find that the majority of initially supplied oxygen goes into CO, while the excess carbon primarily goes into C$_2$H$_2$, HCN, and also CO, leading to a perceptible transience of CH$_4$ \reply{(see Figs.~\ref{fig:ob_X-ray} and \ref{fig:ob_UV}). Lastly, we find that the initial supply of CH$_4$ vs C does not have a significant effect on the equilibrium abundance of most hydrocarbons, with the exception of C$_2$H$_4$, which still gets briefly produced in our C/O $>$ 1 models before being depleted in our previously stated carbon sinks.}

This work serves as an initial attempt to understand the chemical budget and evolution in the cosmic/X-ray \reply{and UV-}powered chemistry limit within the terrestrial planet-forming region in the wake of new constraints provided by the influx of JWST observations. These observations directly call for deeper and more thorough tests of initial chemical conditions to understand the production of hydrocarbons, and ultimately, the methods of carbon grain destruction. This work further has the potential to act as a testbed of the physical and chemical conditions being recovered from JWST spectra, allowing us to benchmark different environments and systems.

\begin{acknowledgments}

F.A. is funded by the European Union (ERC, UNVEIL, 101076613). Views and opinions expressed are however those of the author(s) only and do not necessarily reflect those of the European Union or the European Research Council. Neither the European Union nor the granting authority can be held responsible for them.

\end{acknowledgments}

%

\vspace{5mm}





\appendix
\section{Appendix A}\label{appendix:a}
\begin{figure*}[ht!]
\centering
\includegraphics[width=1\columnwidth]{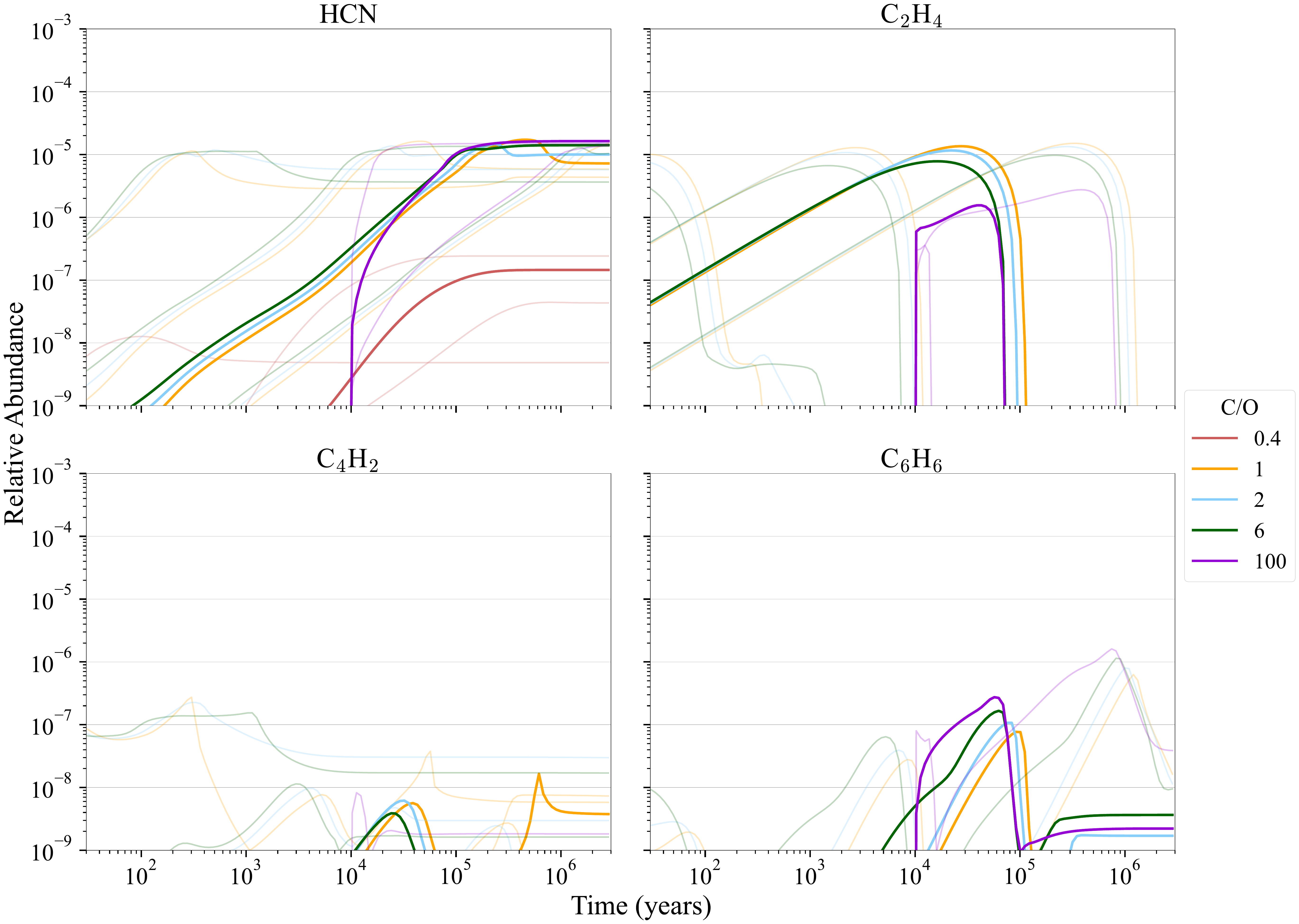}
\caption{\reply{Omnibus plots of HCN, C$_2$H$_4$, C$_4$H$_2$, and C$_6$H$_6$ C showing the relative abundance of these species in the X-ray dominated case, sorted by initial C/O ratio, done in the same manner as Fig.~\ref{fig:ob_X-ray}}}
\label{fig:other_hcarb_xray}
\end{figure*}

\begin{figure*}[h!]b
\centering
\includegraphics[width=1\columnwidth]{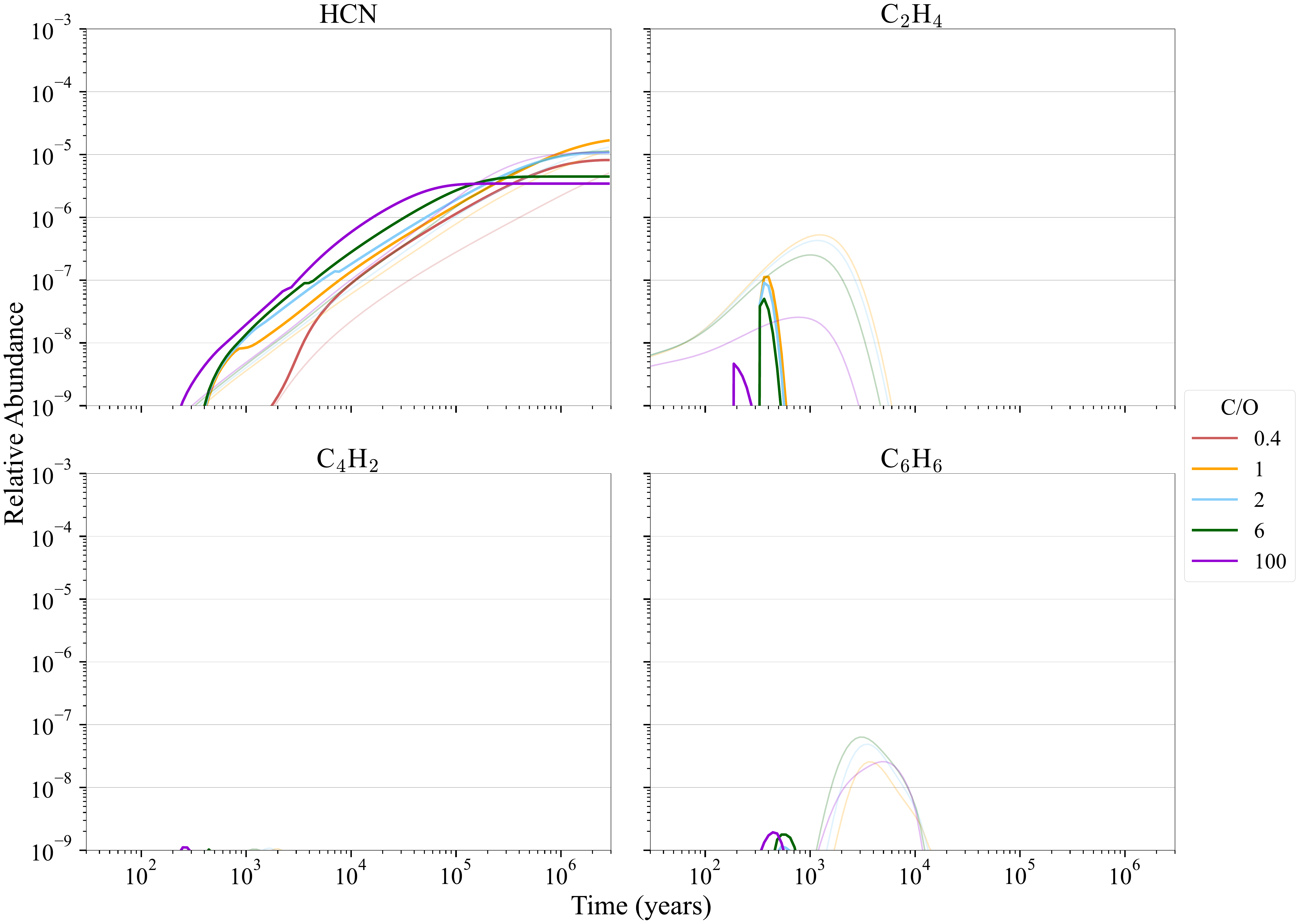}
\caption{\reply{Omnibus plots of HCN, C$_2$H$_4$, C$_4$H$_2$, and C$_6$H$_6$ showing the relative abundance of these species in the UV dominated case, sorted by initial C/O ratio, done in the same manner as Fig.~\ref{fig:ob_X-ray}}}
\label{fig:other_hcarb_uv}
\end{figure*}

\clearpage

\bibliography{sample631}{}

\begin{thebibliography}{}
\expandafter\ifx\csname natexlab\endcsname\relax\def\natexlab#1{#1}\fi
\providecommand{\url}[1]{\href{#1}{#1}}
\providecommand{\dodoi}[1]{doi:~\href{http://doi.org/#1}{\nolinkurl{#1}}}
\providecommand{\doeprint}[1]{\href{http://ascl.net/#1}{\nolinkurl{http://ascl.net/#1}}}
\providecommand{\doarXiv}[1]{\href{https://arxiv.org/abs/#1}{\nolinkurl{https://arxiv.org/abs/#1}}}

\bibitem[{{Ag{\'u}ndez} {et~al.}(2010){Ag{\'u}ndez}, {Goicoechea}, {Cernicharo}, {Faure}, \& {Roueff}}]{Agundez..2010}
{Ag{\'u}ndez}, M., {Goicoechea}, J.~R., {Cernicharo}, J., {Faure}, A., \& {Roueff}, E. 2010, \apj, 713, 662, \dodoi{10.1088/0004-637X/713/1/662}

\bibitem[{{Aikawa} \& {Herbst}(1999)}]{Aikawa1999}
{Aikawa}, Y., \& {Herbst}, E. 1999, \aap, 351, 233

\bibitem[{{Aikawa} {et~al.}(2021){Aikawa}, {Cataldi}, {Yamato}, {Zhang}, {Booth}, {Furuya}, {Andrews}, {Bae}, {Bergin}, {Bergner}, {Bosman}, {Cleeves}, {Czekala}, {Guzm{\'a}n}, {Huang}, {Ilee}, {Law}, {Le Gal}, {Loomis}, {M{\'e}nard}, {Nomura}, {{\"O}berg}, {Qi}, {Schwarz}, {Teague}, {Tsukagoshi}, {Walsh}, \& {Wilner}}]{Aikawa2021}
{Aikawa}, Y., {Cataldi}, G., {Yamato}, Y., {et~al.} 2021, \apjs, 257, 13, \dodoi{10.3847/1538-4365/ac143c}

\bibitem[{{Alarc{\'o}n} {et~al.}(2020){Alarc{\'o}n}, {Teague}, {Zhang}, {Bergin}, \& {Barraza-Alfaro}}]{Alarcon2020}
{Alarc{\'o}n}, F., {Teague}, R., {Zhang}, K., {Bergin}, E.~A., \& {Barraza-Alfaro}, M. 2020, \apj, 905, 68, \dodoi{10.3847/1538-4357/abc1d6}

\bibitem[{{Alarc{\'o}n} {et~al.}(2021){Alarc{\'o}n}, {Bosman}, {Bergin}, {Zhang}, {Teague}, {Bae}, {Aikawa}, {Andrews}, {Booth}, {Calahan}, {Cataldi}, {Czekala}, {Huang}, {Ilee}, {Law}, {Le Gal}, {Liu}, {Long}, {Loomis}, {M{\'e}nard}, {{\"O}berg}, {Schwarz}, {van't Hoff}, {Walsh}, \& {Wilner}}]{Alarcon2021}
{Alarc{\'o}n}, F., {Bosman}, A.~D., {Bergin}, E.~A., {et~al.} 2021, \apjs, 257, 8, \dodoi{10.3847/1538-4365/ac22ae}

\bibitem[{{Anderson} {et~al.}(2017){Anderson}, {Bergin}, {Blake}, {Ciesla}, {Visser}, \& {Lee}}]{Anderson..2017}
{Anderson}, D.~E., {Bergin}, E.~A., {Blake}, G.~A., {et~al.} 2017, \apj, 845, 13, \dodoi{10.3847/1538-4357/aa7da1}

\bibitem[{{Anderson} {et~al.}(2021){Anderson}, {Blake}, {Cleeves}, {Bergin}, {Zhang}, {Schwarz}, {Salyk}, \& {Bosman}}]{Anderson21}
{Anderson}, D.~E., {Blake}, G.~A., {Cleeves}, L.~I., {et~al.} 2021, \apj, 909, 55, \dodoi{10.3847/1538-4357/abd9c1}

\bibitem[{{Arabhavi} {et~al.}(2024){Arabhavi}, {Kamp}, {Henning}, {van Dishoeck}, {Christiaens}, {Gasman}, {Perrin}, {G{\"u}del}, {Tabone}, {Kanwar}, {Waters}, {Pascucci}, {Samland}, {Perotti}, {Bettoni}, {Grant}, {Lagage}, {Ray}, {Vandenbussche}, {Absil}, {Argyriou}, {Barrado}, {Boccaletti}, {Bouwman}, {Caratti o Garatti}, {Glauser}, {Lahuis}, {Mueller}, {Olofsson}, {Pantin}, {Scheithauer}, {Morales-Calder{\'o}n}, {Franceschi}, {Jang}, {Pawellek}, {Rodgers-Lee}, {Schreiber}, {Schwarz}, {Temmink}, {Vlasblom}, {Wright}, {Colina}, \& {{\"O}stlin}}]{Arabhavi2024}
{Arabhavi}, A.~M., {Kamp}, I., {Henning}, T., {et~al.} 2024, Science, 384, 1086, \dodoi{10.1126/science.adi8147}

\bibitem[{{Bergin} {et~al.}(2007){Bergin}, {Aikawa}, {Blake}, \& {van Dishoeck}}]{Bergin2007_ppv}
{Bergin}, E.~A., {Aikawa}, Y., {Blake}, G.~A., \& {van Dishoeck}, E.~F. 2007, in Protostars and Planets V, ed. B.~{Reipurth}, D.~{Jewitt}, \& K.~{Keil}, 751, \dodoi{10.48550/arXiv.astro-ph/0603358}

\bibitem[{{Bergin} {et~al.}(2015){Bergin}, {Blake}, {Ciesla}, {Hirschmann}, \& {Li}}]{Bergin2015}
{Bergin}, E.~A., {Blake}, G.~A., {Ciesla}, F., {Hirschmann}, M.~M., \& {Li}, J. 2015, Proceedings of the National Academy of Science, 112, 8965, \dodoi{10.1073/pnas.1500954112}

\bibitem[{{Bergin} {et~al.}(1998){Bergin}, {Melnick}, \& {Neufeld}}]{Bergin98}
{Bergin}, E.~A., {Melnick}, G.~J., \& {Neufeld}, D.~A. 1998, \apj, 499, 777, \dodoi{10.1086/305656}

\bibitem[{Bern{\'e} {et~al.}(2023)Bern{\'e}, Martin-Drumel, Schroetter, Goicoechea, Jacovella, Gans, Dartois, Coudert, Bergin, Alarcon, Cami, Roueff, Black, Asvany, Habart, Peeters, Canin, Trahin, Joblin, Schlemmer, Thorwirth, Cernicharo, Gerin, Tielens, Zannese, Abergel, Bernard-Salas, Boersma, Bron, Chown, Cuadrado, Dicken, Elyajouri, Fuente, Gordon, Issa, Kannavou, Khan, Lacinbala, Languignon, Le~Gal, Maragkoudakis, Meshaka, Okada, Onaka, Pasquini, Pound, Robberto, R{\"o}llig, Schirmer, Sidhu, Tabone, Van De~Putte, Vicente, \& Wolfire}]{Berne2023}
Bern{\'e}, O., Martin-Drumel, M.-A., Schroetter, I., {et~al.} 2023, Nature, \dodoi{10.1038/s41586-023-06307-x}

\bibitem[{{Bethell} \& {Bergin}(2011)}]{Bethell2011}
{Bethell}, T.~J., \& {Bergin}, E.~A. 2011, \apj, 740, 7, \dodoi{10.1088/0004-637X/740/1/7}

\bibitem[{{Binkert} \& {Birnstiel}(2023)}]{Binkert2023}
{Binkert}, F., \& {Birnstiel}, T. 2023, \mnras, 520, 2055, \dodoi{10.1093/mnras/stad182}

\bibitem[{{Birnstiel} {et~al.}(2016){Birnstiel}, {Fang}, \& {Johansen}}]{Birnstiel..2016}
{Birnstiel}, T., {Fang}, M., \& {Johansen}, A. 2016, \ssr, 205, 41, \dodoi{10.1007/s11214-016-0256-1}

\bibitem[{{Bosman} {et~al.}(2022){Bosman}, {Bergin}, {Calahan}, \& {Duval}}]{Bosman2022}
{Bosman}, A.~D., {Bergin}, E.~A., {Calahan}, J., \& {Duval}, S.~E. 2022, \apjl, 930, L26, \dodoi{10.3847/2041-8213/ac66ce}

\bibitem[{{Broome} {et~al.}(2023){Broome}, {Kama}, {Booth}, \& {Shorttle}}]{Broome..2013}
{Broome}, M., {Kama}, M., {Booth}, R., \& {Shorttle}, O. 2023, \mnras, 522, 3378, \dodoi{10.1093/mnras/stad1159}

\bibitem[{{Carr} \& {Najita}(2008)}]{Carr2008}
{Carr}, J.~S., \& {Najita}, J.~R. 2008, Science, 319, 1504, \dodoi{10.1126/science.1153807}

\bibitem[{{Chaabouni} {et~al.}(2012){Chaabouni}, {Bergeron}, {Baouche}, {Dulieu}, {Matar}, {Congiu}, {Gavilan}, \& {Lemaire}}]{Chaabouni..2012}
{Chaabouni}, H., {Bergeron}, H., {Baouche}, S., {et~al.} 2012, \aap, 538, A128, \dodoi{10.1051/0004-6361/201117409}

\bibitem[{{Chiar} {et~al.}(2013){Chiar}, {Tielens}, {Adamson}, \& {Ricca}}]{Chiar13}
{Chiar}, J.~E., {Tielens}, A.~G.~G.~M., {Adamson}, A.~J., \& {Ricca}, A. 2013, \apj, 770, 78, \dodoi{10.1088/0004-637X/770/1/78}

\bibitem[{{Cleeves} {et~al.}(2015){Cleeves}, {Bergin}, {Qi}, {Adams}, \& {{\"O}berg}}]{Cleeves2015}
{Cleeves}, L.~I., {Bergin}, E.~A., {Qi}, C., {Adams}, F.~C., \& {{\"O}berg}, K.~I. 2015, \apj, 799, 204, \dodoi{10.1088/0004-637X/799/2/204}

\bibitem[{{Du} \& {Bergin}(2014)}]{Du..2014}
{Du}, F., \& {Bergin}, E.~A. 2014, \apj, 792, 2, \dodoi{10.1088/0004-637X/792/1/2}

\bibitem[{{Duval} {et~al.}(2022){Duval}, {Bosman}, \& {Bergin}}]{Duval22}
{Duval}, S.~E., {Bosman}, A.~D., \& {Bergin}, E.~A. 2022, \apjl, 934, L25, \dodoi{10.3847/2041-8213/ac822b}

\bibitem[{{Facchini} {et~al.}(2018){Facchini}, {Pinilla}, {van Dishoeck}, \& {de Juan Ovelar}}]{Facchini..2018}
{Facchini}, S., {Pinilla}, P., {van Dishoeck}, E.~F., \& {de Juan Ovelar}, M. 2018, \aap, 612, A104, \dodoi{10.1051/0004-6361/201731390}

\bibitem[{{Fomenkova}(1999)}]{Fomenkova1999}
{Fomenkova}, M.~N. 1999, \ssr, 90, 109, \dodoi{10.1023/A:1005237828783}

\bibitem[{{Gail}(2002)}]{Gail2002}
{Gail}, H.~P. 2002, \aap, 390, 253, \dodoi{10.1051/0004-6361:20020614}

\bibitem[{{Gail} \& {Trieloff}(2017)}]{Gail2017}
{Gail}, H.-P., \& {Trieloff}, M. 2017, \aap, 606, A16, \dodoi{10.1051/0004-6361/201730480}

\bibitem[{{Garrod} {et~al.}(2008){Garrod}, {Widicus Weaver}, \& {Herbst}}]{Garrod2008}
{Garrod}, R.~T., {Widicus Weaver}, S.~L., \& {Herbst}, E. 2008, \apj, 682, 283, \dodoi{10.1086/588035}

\bibitem[{{Gasman} {et~al.}(2023){Gasman}, {van Dishoeck}, {Grant}, {Temmink}, {Tabone}, {Henning}, {Kamp}, {G{\"u}del}, {Lagage}, {Perotti}, {Christiaens}, {Samland}, {Arabhavi}, {Argyriou}, {Abergel}, {Absil}, {Barrado}, {Boccaletti}, {Bouwman}, {Garatti}, {Geers}, {Glauser}, {Guadarrama}, {Jang}, {Kanwar}, {Lahuis}, {Morales-Calder{\'o}n}, {Mueller}, {Nehm{\'e}}, {Olofsson}, {Pantin}, {Pawellek}, {Ray}, {Rodgers-Lee}, {Scheithauer}, {Schreiber}, {Schwarz}, {Vandenbussche}, {Vlasblom}, {Waters}, {Wright}, {Colina}, {Greve}, \& {{\"O}stlin}}]{Gasman2023}
{Gasman}, D., {van Dishoeck}, E.~F., {Grant}, S.~L., {et~al.} 2023, arXiv e-prints, arXiv:2307.09301, \dodoi{10.48550/arXiv.2307.09301}

\bibitem[{{Glassgold} {et~al.}(1997){Glassgold}, {Najita}, \& {Igea}}]{Glassgold1997}
{Glassgold}, A.~E., {Najita}, J., \& {Igea}, J. 1997, \apj, 480, 344, \dodoi{10.1086/303952}

\bibitem[{{Gorti} {et~al.}(2011){Gorti}, {Hollenbach}, {Najita}, \& {Pascucci}}]{Gorti2011}
{Gorti}, U., {Hollenbach}, D., {Najita}, J., \& {Pascucci}, I. 2011, \apj, 735, 90, \dodoi{10.1088/0004-637X/735/2/90}

\bibitem[{{Grant} {et~al.}(2023){Grant}, {van Dishoeck}, {Tabone}, {Gasman}, {Henning}, {Kamp}, {G{\"u}del}, {Lagage}, {Bettoni}, {Perotti}, {Christiaens}, {Samland}, {Arabhavi}, {Argyriou}, {Abergel}, {Absil}, {Barrado}, {Boccaletti}, {Bouwman}, {o Garatti}, {Geers}, {Glauser}, {Guadarrama}, {Jang}, {Kanwar}, {Lahuis}, {Morales-Calder{\'o}n}, {Mueller}, {Nehm{\'e}}, {Olofsson}, {Pantin}, {Pawellek}, {Ray}, {Rodgers-Lee}, {Scheithauer}, {Schreiber}, {Schwarz}, {Temmink}, {Vandenbussche}, {Vlasblom}, {Waters}, {Wright}, {Colina}, {Greve}, {Justannont}, \& {{\"O}stlin}}]{Grant..2023}
{Grant}, S.~L., {van Dishoeck}, E.~F., {Tabone}, B., {et~al.} 2023, \apjl, 947, L6, \dodoi{10.3847/2041-8213/acc44b}

\bibitem[{{Hasegawa} {et~al.}(1992){Hasegawa}, {Herbst}, \& {Leung}}]{Hasegawa..92}
{Hasegawa}, T.~I., {Herbst}, E., \& {Leung}, C.~M. 1992, \apjs, 82, 167, \dodoi{10.1086/191713}

\bibitem[{{Herbst}(2021)}]{Herbst21}
{Herbst}, E. 2021, Frontiers in Astronomy and Space Sciences, 8, 207, \dodoi{10.3389/fspas.2021.776942}

\bibitem[{{Jones} {et~al.}(2013){Jones}, {Fanciullo}, {K{\"o}hler}, {Verstraete}, {Guillet}, {Bocchio}, \& {Ysard}}]{Jones13}
{Jones}, A.~P., {Fanciullo}, L., {K{\"o}hler}, M., {et~al.} 2013, \aap, 558, A62, \dodoi{10.1051/0004-6361/201321686}

\bibitem[{{Kamp} {et~al.}(2010){Kamp}, {Tilling}, {Woitke}, {Thi}, \& {Hogerheijde}}]{Kamp2010}
{Kamp}, I., {Tilling}, I., {Woitke}, P., {Thi}, W.~F., \& {Hogerheijde}, M. 2010, \aap, 510, A18, \dodoi{10.1051/0004-6361/200913076}

\bibitem[{{Kamp} {et~al.}(2023){Kamp}, {Henning}, {Arabhavi}, {Bettoni}, {Christiaens}, {Gasman}, {Grant}, {Morales-Calder{\'o}n}, {Tabone}, {Abergel}, {Absil}, {Argyriou}, {Barrado}, {Boccaletti}, {Bouwman}, {Caratti o Garatti}, {van Dishoeck}, {Geers}, {Glauser}, {G{\"u}del}, {Guadarrama}, {Jang}, {Kanwar}, {Lagage}, {Lahuis}, {Mueller}, {Nehm{\'e}}, {Olofsson}, {Pantin}, {Pawellek}, {Perotti}, {Ray}, {Rodgers-Lee}, {Samland}, {Scheithauer}, {Schreiber}, {Schwarz}, {Temmink}, {Vandenbussche}, {Vlasblom}, {Waelkens}, {Waters}, \& {Wright}}]{Kamp2023}
{Kamp}, I., {Henning}, T., {Arabhavi}, A.~M., {et~al.} 2023, Faraday Discussions, 245, 112, \dodoi{10.1039/D3FD00013C}

\bibitem[{{Kanwar} {et~al.}(2023){Kanwar}, {Kamp}, {Woitke}, {Rab}, {Thi}, \& {Min}}]{Kanwar2023}
{Kanwar}, J., {Kamp}, I., {Woitke}, P., {et~al.} 2023, arXiv e-prints, arXiv:2310.04505, \dodoi{10.48550/arXiv.2310.04505}

\bibitem[{{Kanwar} {et~al.}(2024){Kanwar}, {Kamp}, {Jang}, {Waters}, {van Dishoeck}, {Christiaens}, {Arabhavi}, {Henning}, {G{\"u}del}, {Woitke}, {Absil}, {Barrado}, {Garatti}, {Glauser}, {Lahuis}, {Scheithauer}, {Vandenbussche}, {Gasman}, {Grant}, {Kurtovic}, {Perotti}, {Tabone}, \& {Temmink}}]{Kanwar2024}
{Kanwar}, J., {Kamp}, I., {Jang}, H., {et~al.} 2024, A\&A, in press, \dodoi{10.48550/arXiv.2407.14362}

\bibitem[{{Kaufman} \& {Neufeld}(1996)}]{Kaufman1996}
{Kaufman}, M.~J., \& {Neufeld}, D.~A. 1996, \apj, 456, 611, \dodoi{10.1086/176683}

\bibitem[{{Klarmann} {et~al.}(2018){Klarmann}, {Ormel}, \& {Dominik}}]{Klarmann2018}
{Klarmann}, L., {Ormel}, C.~W., \& {Dominik}, C. 2018, \aap, 618, L1, \dodoi{10.1051/0004-6361/201833719}

\bibitem[{{Lee} {et~al.}(2010){Lee}, {Bergin}, \& {Nomura}}]{Lee2010}
{Lee}, J.-E., {Bergin}, E.~A., \& {Nomura}, H. 2010, \apjl, 710, L21, \dodoi{10.1088/2041-8205/710/1/L21}

\bibitem[{{Li} {et~al.}(2021){Li}, {Bergin}, {Blake}, {Ciesla}, \& {Hirschmann}}]{Li2021}
{Li}, J., {Bergin}, E.~A., {Blake}, G.~A., {Ciesla}, F.~J., \& {Hirschmann}, M.~M. 2021, Science Advances, 7, eabd3632, \dodoi{10.1126/sciadv.abd3632}

\bibitem[{{Mah} {et~al.}(2023){Mah}, {Bitsch}, {Pascucci}, \& {Henning}}]{Mah2023}
{Mah}, J., {Bitsch}, B., {Pascucci}, I., \& {Henning}, T. 2023, \aap, 677, L7, \dodoi{10.1051/0004-6361/202347169}

\bibitem[{{McElroy} {et~al.}(2013){McElroy}, {Walsh}, {Markwick}, {Cordiner}, {Smith}, \& {Millar}}]{McElroy2013}
{McElroy}, D., {Walsh}, C., {Markwick}, A.~J., {et~al.} 2013, \aap, 550, A36, \dodoi{10.1051/0004-6361/201220465}

\bibitem[{{Meijerink} {et~al.}(2009){Meijerink}, {Pontoppidan}, {Blake}, {Poelman}, \& {Dullemond}}]{Meijerink2009}
{Meijerink}, R., {Pontoppidan}, K.~M., {Blake}, G.~A., {Poelman}, D.~R., \& {Dullemond}, C.~P. 2009, \apj, 704, 1471, \dodoi{10.1088/0004-637X/704/2/1471}

\bibitem[{{Millar} {et~al.}(2024){Millar}, {Walsh}, {Van de Sande}, \& {Markwick}}]{Millar2024}
{Millar}, T.~J., {Walsh}, C., {Van de Sande}, M., \& {Markwick}, A.~J. 2024, \aap, 682, A109, \dodoi{10.1051/0004-6361/202346908}

\bibitem[{{Mishra} \& {Li}(2015)}]{Mishra05}
{Mishra}, A., \& {Li}, A. 2015, \apj, 809, 120, \dodoi{10.1088/0004-637X/809/2/120}

\bibitem[{{Najita} {et~al.}(2011){Najita}, {{\'A}d{\'a}mkovics}, \& {Glassgold}}]{Najita2011}
{Najita}, J.~R., {{\'A}d{\'a}mkovics}, M., \& {Glassgold}, A.~E. 2011, \apj, 743, 147, \dodoi{10.1088/0004-637X/743/2/147}

\bibitem[{{Nieva} \& {Przybilla}(2012)}]{Nieva..2012}
{Nieva}, M.~F., \& {Przybilla}, N. 2012, \aap, 539, A143, \dodoi{10.1051/0004-6361/201118158}

\bibitem[{{Nomura} {et~al.}(2007){Nomura}, {Aikawa}, {Tsujimoto}, {Nakagawa}, \& {Millar}}]{Nomura2007}
{Nomura}, H., {Aikawa}, Y., {Tsujimoto}, M., {Nakagawa}, Y., \& {Millar}, T.~J. 2007, \apj, 661, 334, \dodoi{10.1086/513419}

\bibitem[{{Nomura} \& {Millar}(2005)}]{Nomura2005}
{Nomura}, H., \& {Millar}, T.~J. 2005, \aap, 438, 923, \dodoi{10.1051/0004-6361:20052809}

\bibitem[{{Pascucci} {et~al.}(2009){Pascucci}, {Apai}, {Luhman}, {Henning}, {Bouwman}, {Meyer}, {Lahuis}, \& {Natta}}]{Pascucci2009}
{Pascucci}, I., {Apai}, D., {Luhman}, K., {et~al.} 2009, \apj, 696, 143, \dodoi{10.1088/0004-637X/696/1/143}

\bibitem[{{Pascucci} {et~al.}(2013){Pascucci}, {Herczeg}, {Carr}, \& {Bruderer}}]{Pascucci2013}
{Pascucci}, I., {Herczeg}, G., {Carr}, J.~S., \& {Bruderer}, S. 2013, \apj, 779, 178, \dodoi{10.1088/0004-637X/779/2/178}

\bibitem[{{Pontoppidan} {et~al.}(2010){Pontoppidan}, {Salyk}, {Blake}, {Meijerink}, {Carr}, \& {Najita}}]{Pontoppidan2009}
{Pontoppidan}, K.~M., {Salyk}, C., {Blake}, G.~A., {et~al.} 2010, \apj, 720, 887, \dodoi{10.1088/0004-637X/720/1/887}

\bibitem[{{Preibisch} {et~al.}(2005){Preibisch}, {Kim}, {Favata}, {Feigelson}, {Flaccomio}, {Getman}, {Micela}, {Sciortino}, {Stassun}, {Stelzer}, \& {Zinnecker}}]{Preibisch2005}
{Preibisch}, T., {Kim}, Y.-C., {Favata}, F., {et~al.} 2005, \apjs, 160, 401, \dodoi{10.1086/432891}

\bibitem[{Rackauckas \& Nie(2017)}]{Rackauckas}
Rackauckas, C., \& Nie, Q. 2017, The Journal of Open Research Software, 5, \dodoi{10.5334/jors.151}

\bibitem[{{Ramirez-Tannus} {et~al.}(2023){Ramirez-Tannus}, {Bik}, {Cuijpers}, {Waters}, {Goppl}, {Henning}, {Kamp}, {Preibisch}, {Getman}, {Chaparro}, {Cuartas-Restrepo}, {de Koter}, {Feigelson}, {Grant}, {Haworth}, {Hern{\'a}ndez}, {Kuhn}, {Perotti}, {Povich}, {Reiter}, {Roccatagliata}, {Sabbi}, {Tabone}, {Winter}, {McLeod}, {van Boekel}, \& {van Terwisga}}]{Tannus2023}
{Ramirez-Tannus}, M.~C., {Bik}, A., {Cuijpers}, L., {et~al.} 2023, arXiv e-prints, arXiv:2310.11074, \dodoi{10.48550/arXiv.2310.11074}

\bibitem[{{Rubin} {et~al.}(2019){Rubin}, {Altwegg}, {Balsiger}, {Berthelier}, {Combi}, {De Keyser}, {Drozdovskaya}, {Fiethe}, {Fuselier}, {Gasc}, {Gombosi}, {H{\"a}nni}, {Hansen}, {Mall}, {R{\`e}me}, {Schroeder}, {Schuhmann}, {S{\'e}mon}, {Waite}, {Wampfler}, \& {Wurz}}]{Rubin2019}
{Rubin}, M., {Altwegg}, K., {Balsiger}, H., {et~al.} 2019, \mnras, 489, 594, \dodoi{10.1093/mnras/stz2086}

\bibitem[{{Salyk} {et~al.}(2011){Salyk}, {Pontoppidan}, {Blake}, {Najita}, \& {Carr}}]{Salyk2011}
{Salyk}, C., {Pontoppidan}, K.~M., {Blake}, G.~A., {Najita}, J.~R., \& {Carr}, J.~S. 2011, \apj, 731, 130, \dodoi{10.1088/0004-637X/731/2/130}

\bibitem[{{Seifert} {et~al.}(2021){Seifert}, {Cleeves}, {Adams}, \& {Li}}]{Seifert2021}
{Seifert}, R.~A., {Cleeves}, L.~I., {Adams}, F.~C., \& {Li}, Z.-Y. 2021, \apj, 912, 136, \dodoi{10.3847/1538-4357/abf09a}

\bibitem[{{Tabone} {et~al.}(2023){Tabone}, {Bettoni}, {van Dishoeck}, {Arabhavi}, {Grant}, {Gasman}, {Henning}, {Kamp}, {G{\"u}del}, {Lagage}, {Ray}, {Vandenbussche}, {Abergel}, {Absil}, {Argyriou}, {Barrado}, {Boccaletti}, {Bouwman}, {Caratti o Garatti}, {Geers}, {Glauser}, {Justannont}, {Lahuis}, {Mueller}, {Nehm{\'e}}, {Olofsson}, {Pantin}, {Scheithauer}, {Waelkens}, {Waters}, {Black}, {Christiaens}, {Guadarrama}, {Morales-Calder{\'o}n}, {Jang}, {Kanwar}, {Pawellek}, {Perotti}, {Perrin}, {Rodgers-Lee}, {Samland}, {Schreiber}, {Schwarz}, {Colina}, {{\"O}stlin}, \& {Wright}}]{Tabone2023}
{Tabone}, B., {Bettoni}, G., {van Dishoeck}, E.~F., {et~al.} 2023, Nature Astronomy, 7, 805, \dodoi{10.1038/s41550-023-01965-3}

\bibitem[{{van der Marel} {et~al.}(2018){van der Marel}, {Williams}, \& {Bruderer}}]{Nienke..2018}
{van der Marel}, N., {Williams}, J.~P., \& {Bruderer}, S. 2018, \apjl, 867, L14, \dodoi{10.3847/2041-8213/aae88e}

\bibitem[{{van Dishoeck} {et~al.}(2023){van Dishoeck}, {Grant}, {Tabone}, {van Gelder}, {Francis}, {Tychoniec}, {Bettoni}, {Arabhavi}, {Gasman}, {Nazari}, {Vlasblom}, {Kavanagh}, {Christiaens}, {Klaassen}, {Beuther}, {Henning}, \& {Kamp}}]{Dishoeck2023}
{van Dishoeck}, E.~F., {Grant}, S., {Tabone}, B., {et~al.} 2023, Faraday Discussions, 245, 52, \dodoi{10.1039/D3FD00010A}

\bibitem[{{Wagner} \& {Graff}(1987)}]{Wagner1987}
{Wagner}, A.~F., \& {Graff}, M.~M. 1987, \apj, 317, 423, \dodoi{10.1086/165287}

\bibitem[{{Wei} {et~al.}(2019){Wei}, {Nomura}, {Lee}, {Ip}, {Walsh}, \& {Millar}}]{Wei2019}
{Wei}, C.-E., {Nomura}, H., {Lee}, J.-E., {et~al.} 2019, \apj, 870, 129, \dodoi{10.3847/1538-4357/aaf390}

\bibitem[{{Woitke} {et~al.}(2024){Woitke}, {Thi}, {Arabhavi}, {Kamp}, {K{\'o}sp{\'a}l}, \& {{\'A}brah{\'a}m}}]{Woitke2024}
{Woitke}, P., {Thi}, W.~F., {Arabhavi}, A.~M., {et~al.} 2024, \aap, 683, A219, \dodoi{10.1051/0004-6361/202347730}

\bibitem[{{Woodall} {et~al.}(2007){Woodall}, {Ag{\'u}ndez}, {Markwick-Kemper}, \& {Millar}}]{Woodall2007}
{Woodall}, J., {Ag{\'u}ndez}, M., {Markwick-Kemper}, A.~J., \& {Millar}, T.~J. 2007, \aap, 466, 1197, \dodoi{10.1051/0004-6361:20064981}

\end{thebibliography}
\bibliographystyle{aasjournal}



\end{document}